\documentclass[nofootinbib,a4paper,aps,prd,10pt,superscriptaddress,showkeys,twocolumn]{revtex4}

\usepackage{graphicx}
\usepackage{amsfonts}
\usepackage{amssymb} 
\usepackage{amsmath}
\usepackage{hyperref}
\usepackage{natbib}
\usepackage{float}
\usepackage{dcolumn}
\usepackage{bm}
\usepackage{latexsym,color}

\newcommand{\bea}{\begin{eqnarray}}
\newcommand{\eea}{\end{eqnarray}}
\newcommand{\be}{\begin{equation}}
\newcommand{\ee}{\end{equation}}

\begin{document}

\title{Geometric properties versus particle motion in the Fan-Wang spacetime}

\author{Gulnara~\surname{Suliyeva}}
\email[]{g\_suliyeva@mail.ru}
\affiliation{National Nanotechnology Laboratory of Open Type,  Almaty 050040, Kazakhstan.}
\affiliation{Al-Farabi Kazakh National University, Al-Farabi av. 71, Almaty 050040, Kazakhstan.}
\affiliation{Fesenkov Astrophysical Institute, Observatory 23, Almaty 050020, Kazakhstan.}

\author{Kuantay~\surname{Boshkayev}}
\email[]{kuantay@mail.ru}
\affiliation{National Nanotechnology Laboratory of Open Type,  Almaty 050040, Kazakhstan.}
\affiliation{Al-Farabi Kazakh National University, Al-Farabi av. 71, Almaty 050040, Kazakhstan.}
\affiliation{Institute of Nuclear Physics, Ibragimova, 1, Almaty 050032, Kazakhstan.}

\author{Talgar~\surname{Konysbayev}}
\email[] {talgar\_777@mail.ru}
\affiliation{National Nanotechnology Laboratory of Open Type,  Almaty 050040, Kazakhstan.}

\affiliation{Al-Farabi Kazakh National University, Al-Farabi av. 71, Almaty 050040, Kazakhstan.}

\author{Yergali~\surname{Kurmanov}}
\email[]{kurmanov.yergali@kaznu.kz}
\affiliation{National Nanotechnology Laboratory of Open Type,  Almaty 050040, Kazakhstan.}

\affiliation{Al-Farabi Kazakh National University, Al-Farabi av. 71,  Almaty 050040, Kazakhstan.}

\author{Orlando~\surname{Luongo}}
\email[]{orlando.luongo@unicam.it}
\affiliation{Al-Farabi Kazakh National University, Al-Farabi av. 71, Almaty 050040, Kazakhstan.}
\affiliation{Universit\`a di Camerino, Divisione di Fisica, Via Madonna delle carceri 9, 62032 Camerino, Italy}
\affiliation{Department of Nanoscale Science and Engineering, University at Albany-SUNY, Albany, New York 12222, USA.}
\affiliation{INAF - Osservatorio Astronomico di Brera, Milano, Italy.}
\affiliation{Istituto Nazionale di Fisica Nucleare (INFN), Sezione di Perugia, Perugia, 06123, Italy.}

\author{Marco~\surname{Muccino}}
\email[]{marco.muccino@lnf.infn.it}
\affiliation{Al-Farabi Kazakh National University, Al-Farabi av. 71, Almaty 050040, Kazakhstan.}
\affiliation{Universit\`a di Camerino, Divisione di Fisica, Via Madonna delle carceri 9, 62032 Camerino, Italy}
\affiliation{ICRANet, Piazza della Repubblica 10, 65122 Pescara, Italy.}

\author{Hernando~\surname{Quevedo}}
\email[]{quevedo@nucleares.unam.mx}
\affiliation{Al-Farabi Kazakh National University, Al-Farabi av. 71, Almaty 050040, Kazakhstan.}
\affiliation{Instituto de Ciencias Nucleares, Universidad Nacional Aut\'onoma de M\'exico, Mexico.}%
\affiliation{Dipartimento di Fisica and Icra, Universit\`a di Roma “La Sapienza”, Roma, Italy.}

\author{Ainur~\surname{Urazalina}}
\email[]{y.a.a.707@mail.ru}
\affiliation{National Nanotechnology Laboratory of Open Type,  Almaty 050040, Kazakhstan.}
\affiliation{Al-Farabi Kazakh National University, Al-Farabi av. 71, Almaty 050040, Kazakhstan.}
\affiliation{Institute of Nuclear Physics, Ibragimova, 1, Almaty 050032, Kazakhstan.}
\date{\today}

\author{Farida~\surname{Belissarova}}
\email[]{farida.belisarova@kaznu.kz}
\affiliation{Al-Farabi Kazakh National University, Al-Farabi av. 71, Almaty 050040, Kazakhstan.}

\author{Anar~\surname{Dalelkhankyzy}}
\email[]{dalelkhankyzy.d@gmail.com}
\affiliation{Kazakh National Women's Teacher Training University, Ayteke Bi, 99, Almaty 050000, Kazakhstan.}

\date{\today}

\begin{abstract}
In this work, we explore general relativistic effects and geometric properties of the Fan-Wang spacetime, one of the simplest regular solutions that can be obtained in nonlinear electrodynamics. In particular, we investigate the motion of test particles, the capture cross-section of neutral massive and massless particles, such as neutrinos and photons, and the gravitational redshift. Additionally, using a perturbative approach, we derive analytical expressions for the perihelion shift and gravitational deflection of massless particles. By identifying the one-parameter corrections to the Schwarzschild spacetime, induced by the magnetic charge contained in the Fan-Wang metric, we show that this spacetime can be falsified, since it modifies classical general relativity predictions even at the local level. Moreover, we argue that these modifications could be experimentally tested with advanced observational instrumentation.
\end{abstract}


\maketitle

 \section{Introduction}
 \label{sec:Intro}

Black holes (BHs) are the best-suited natural laboratories for exploring extreme gravitational fields and testing the limits of general relativity (GR), as confirmed by gravitational wave astronomy \cite{2016PhRvL.116v1101A} and the Event Horizon Telescope’s high-resolution imaging of the shadows of the super massive BHs of M87$^*$ \cite{2019ApJ...875L...1E} and Sgr A$^*$ \cite{2022ApJ...930L..12E}.

However, classical BH solutions --- such as the Schwarzschild \cite{1916AbhKP1916..189S}, the Reissner--Nordstr\"{o}m \cite{1916AnP...355..106R,1918KNAB...20.1238N}, and the Kerr \cite{1963PhRvL..11..237K} metrics --- possess singularities, posing fundamental challenges for gravitational physics. To overcome this issue, regular black hole (RBH) models have been proposed, introducing modifications to the geometry, often through the coupling with nonlinear electrodynamics (NED). These models ensure that spacetime curvature remains finite everywhere while preserving key physical properties.
In this context, Bardeen \cite{1968qtr..conf...87B} introduced the first static, spherically symmetric BH solution with a regular core, which remains asymptotically flat at large distances. Although not initially derived as an exact solution to Einstein’s field equations, Ayón-Beato and García \cite{1998PhRvL..80.5056A,2000PhLB..493..149A} and Bronnikov \cite{2001PhRvD..63d4005B} later demonstrated that the Bardeen BH could be realized as a solution to Einstein’s equations when coupled with a NED source corresponding to a magnetic monopole.

Following Bardeen’s pioneering work, other RBH models have been developed. For example, Dymnikova \cite{1992GReGr..24..235D} proposed a model that coincides with Schwarzschild spacetime at large distances while transitioning to de Sitter spacetime near the center. Hayward \cite{PhysRevLett.96.031103} introduced another static, spherically symmetric RBH model as a potential resolution to the BH information loss paradox. Among these, the Fan-Wang solution \cite{2016PhRvD..94l4027F} stands out for its versatility, offering a framework for constructing a broad class of spherically symmetric magnetic solutions. For comprehensive reviews on RBH solutions, their stability, and thermodynamics, see Refs.~\cite{2022JHEP...11..108M,2023IJTP...62..202L,bronnikov2023regular}.

It is worth noting that current astrophysical observations do not provide conclusive evidence for the existence of a monopole magnetic field or, by extension, a magnetic charge. However, advancements in observational and experimental techniques offer hope for detecting magnetic charges in nonlinear electromagnetic fields. One promising approach involves analyzing particle dynamics, which could lead to new observational tests of nonlinear fields and their associated charges.

In the framework of NED, photons do not follow the null geodesics of the RBH spacetime but instead propagate along null geodesics of an effective geometry \cite{PhysRevD.61.045001,2019ApJ...874...12S,2019ApJ...887..145S,2019EPJC...79...44S,2019EPJC...79..988S,2023PhRvD.108h4029D}.
For this reason, the dynamics of massless particles and photons in the Fan-Wang spacetime is governed by different equations of motion. 
Extensive research has explored particle dynamics in the vicinity of various RBH models \cite{2014Ap&SS.352..769A,2015JMP....56c2501G,2015IJMPD..2450020S,2015Ap&SS.357...41T,2019EPJC...79..778V,2019CaJPh..97...58H,2020PhRvD.101j4045R,2020PhRvD.102j4062N,2020AnPhy.41868194G,2021arXiv210706085U,2022IJMPD..3150032R,2023PhRvD.107h4003I,2020PhRvD.102j4033P}.

Post-Newtonian (PN) approximations have been widely used to study particle dynamics, providing analytical solutions for the two-body problem, including first- and higher-order PN corrections \cite{1972gcpa.book.....W,1972rcm..book.....B,1987CeMec..40...77S,1993tegp.book.....W,1994ApJ...427..951K,1988NCimB.101..127D,narlikar2010introduction,2017PhRvD..96d4011B}. The PN framework offers a more accurate description of gravitational interactions, particularly in cases where obtaining an exact analytical solution to the equations of motion is infeasible. It also plays a crucial role in testing deviations from general relativity (GR), especially in the weak-field regime \cite{2014grav.book.....P}.

Notably, Ref.~\cite{2023GReGr..55..114L} derived the second PN solution for the quasi-Keplerian motion of a test particle in the gravitational field of a Bardeen BH. This solution is expressed in terms of key parameters, including the particle's orbital energy, angular momentum, and the BH’s mass and magnetic charge. The study reveals the impact of the magnetic charge on orbital dynamics, particularly perihelion precession. Interestingly, at the second PN order, the magnetic charge does not influence the orbital period of the test particle.

Remarkably, within the general class of the so-called Fan-Wang solutions, Ref.~\cite{2023PhRvD.107h4003I} investigated the bound orbits of neutral massive and massless particles, including photons. The study demonstrated the existence of both stable and unstable circular orbits for massive particles, as well as the dependence of the innermost stable circular orbit (ISCO) on the charge parameter\footnote{Furthermore, for massless particles and photons, it was shown that stable and unstable circular orbits can exist even in spacetimes with a slight overcharge. Additionally, an analytical expression for the periapsis shift was derived, revealing that the charge introduces a negative correction to the shift in cases of small nonlinearity.}.

In Ref.~\cite{2024PDU....4601566K}, the radiative properties of accretion disks were investigated for the Fan-Wang and Dymnikova spacetimes. By analyzing circular geodesics and employing the Novikov-Thorne thin accretion disk formalism, the differential luminosity, surface temperature, and efficiency of mass-to-radiation conversion were calculated. The results, compared to those for the Schwarzschild spacetime, revealed an increase in energy emission from the disk surface, leading to higher temperatures for certain values of the charge parameters. Notably, the efficiency of mass-to-radiation conversion was found to be highest for the Fan-Wang metric compared to the Dymnikova and Schwarzschild solutions, making this spacetime particularly promising for distinguishing between different BH models using observational data. Additionally, thin accretion disks have been studied in the contexts of Bardeen and Hayward RBHs \cite{akbarieh2024}, as well as their rotating counterparts \cite{2024EPJC...84..230B,kurmanov2024radiative}.

The authors of Ref.~\cite{2019ApJ...887..145S} investigated optical phenomena within the general Fan-Wang BH spacetime. A comparative analysis was conducted between the trajectories of photons and massless neutrino-like particles, which are unaffected by the nonlinear effects of non-Maxwellian electromagnetic fields. Key aspects examined include BH shadows, relative time delays experienced by photons and neutrinos in the gravitational field, and the construction of images for Keplerian disks surrounding these BHs. The results suggest that, for Maxwellian NED BHs, observable optical phenomena could serve as potential signatures of NED effects, which may be detectable by instruments such as GRAVITY or the Event Horizon Telescope. However, these effects remain sufficiently subtle to avoid contradictions with current observational constraints, similar to the behavior observed in the Bardeen BH \cite{2019ApJ...874...12S}.

Motivated by the above findings on the Fan-Wang metric, in this work we analyze some relevant gravitational effects and geometric properties within the aforementioned Fan-Wang spacetime. In particular, we focus on gravitational capture, perihelion shift, gravitational deflection, and gravitational redshift. To this end, we employ a systematic approach based on the weak-field approximation, allowing us to quantify deviations from the Schwarzschild spacetime due to the presence of a magnetic charge. To derive analytical expressions for the perihelion shift and the deflection angle of massless particles and photons, we use a perturbative approach, expanding the equations of motion to incorporate PN corrections. Gravitational capture is examined using a standard methodology that evaluates critical impact parameters distinguishing captured trajectories from scattered ones. These calculations are performed for neutral massive particles, massless particles, and photons moving in the equatorial plane. Finally, gravitational redshift is analyzed by computing the frequency shift of photons emitted near the BH and observed at infinity, enabling us to characterize modifications to general relativity induced by NED effects.

The paper is structured as follows. Section \ref{sec:FW_BH} provides an overview of the main properties of the Fan-Wang spacetime. In Section \ref{sec:geod_motion}, we analyze the geodesic motion of neutral massive and massless particles, as well as photons. Sections \ref{sec:grav_capt}, \ref{sec:perihelion}, \ref{sec:grav_defl}, and \ref{sec:grav_redshift} are dedicated to gravitational capture, perihelion shift, gravitational deflection, and gravitational redshift, respectively. Finally, in Section \ref{sec:concl}, we present our concluding remarks. Throughout the paper, we adopt geometrical units, setting $G=c=1$.

\section{Theoretical set up} \label{sec:FW_BH}

We provide a brief overview of the key characteristics of the Fan-Wang spacetime, which serves as the foundation for understanding the effects examined in this study.

\subsection{Reproducing regular spacetimes from NED}

The RBH solution connected to NED is derived from a modified action \cite{2016PhRvD..94l4027F, PhysRevD.98.028501}
\begin{equation}\label{action}
    \frac{1}{16\pi G}\int d^4x\sqrt{-g}\left(R-\mathcal{L}(\mathcal{F})\right)
\end{equation}
that incorporates NED with the Lagrangian density $\mathcal{L}(\mathcal{F})$, where $\mathcal{F}=F_{\alpha\beta}F^{\alpha\beta}$ represents the scalar of electromagnetic field strength tensor, $R$ is the Ricci scalar, $g$ is the metric determinant, $G$ is the gravitational constant. In the generic RBH spacetime case, the Lagrangian density reads:
\begin{equation}\label{NED_lagr}
    \mathcal{L}(\mathcal{F}) = \frac{4\mu}{\alpha}\frac{\left(\alpha\mathcal{F}\right)^{\frac{\nu+3}{4}}}{\left[1+\left(\alpha\mathcal{F}\right)^{\frac{\nu}{4}}\right]^{\frac{\mu+\nu}{\nu}}},
\end{equation}
where $\mu$ and $\nu$ are dimensionless parameters. The general class of BH spacetimes includes previously known RBH solutions. Specifically, when the parameter $\mu=3$ is chosen to ensure regularity, setting $\nu=2$ yields the Bardeen spacetime \cite{1968qtr..conf...87B}, and $\nu=3$ gives the Hayward spacetime \cite{PhysRevLett.96.031103}. An interesting case arises for $\nu=1$, leading to the Fan-Wang spacetime. The importance of the last case, ($\mu$, $\nu$)=(3, 1), lies in the fact that, in the weak field limit, $\mathcal{F}\to 0$, the theory retains consistency with Maxwell's electrodynamics. 
The parameter $\alpha>0$ has the dimension of length squared and relates to the magnetic charge $q_m$ as
\begin{equation}\label{magn_charge}
    q_m = \frac{1}{4\pi}\int_{S^{2}} F = \frac{l^2}{\sqrt{2\alpha}}, \quad \mathcal F=\frac{2q_m^2}{r^4},
\end{equation}
where

\begin{eqnarray}
    F=F_{\theta\phi}d\theta d\phi, \quad F_{\theta\phi} = \partial_{\phi}A_{\theta}-\partial_{\theta}A_{\phi},\nonumber\\ A=A_{\alpha}dx^{\alpha}=A_{\phi}d\phi=\frac{l^2}{\sqrt{2\alpha}}\cos \theta d\phi,
\end{eqnarray}
and $F_{\theta\phi}=-F_{\phi\theta}$ is non-zero component of field strength tensor $F_{\alpha\beta}$, $A_{\alpha}$ is a vector potential. The charge parameter $l$ is constrained by setting
\begin{equation}
    m_{\text{em}}=l^3/\alpha
\end{equation}
with $m_{\text{em}}$ being the electromagnetically induced gravitational mass. All details on the specified parameters can be found in Refs.~\cite{2016PhRvD..94l4027F, 2018PhRvD..98b8501T}.

The generic static spherically-symmetric spacetime in standard curvature coordinates is given by
\begin{equation}\label{metr_generic}
    ds^2=-f(r)dt^2+\frac{dr^2}{f(r)}+r^2(d\theta^2+\sin^2{\theta}d\phi^2)
\end{equation}
with the lapse function\footnote{This form of a lapse function for RBHs was proposed in \cite{PhysRevD.106.044028} by applying the Damour-Solodukhin prescription.}
\begin{equation}
    f(r)=1-\frac{2M(r)}{r}.
\end{equation}
The mass function $M(r)$ of the Fan-Wang solution takes the form\footnote{We use the metric function given in Ref.~\cite{2022JHEP...11..108M} and not the form that was originally introduced in Ref.~\cite{2016PhRvD..94l4027F}.}
\begin{equation}
    M(r)=\frac{m r^3}{(r+l)^3} 
\end{equation}
and reduces to the Arnowitt-Deser-Misner (ADM) mass $m$ at $r \to \infty$, meaning that the spacetime is asymptotically flat. Moreover, at $l=0$, the Schwarzschild solution is recovered. 

\subsection{Spacetime horizons}

The Fan-Wang RBH horizons are determined by solving equation $g_{tt}=0$, or, equivalently,
\begin{equation}
    (r+l)^3-2mr^2=0,
\end{equation}
whose solution yields two distinct roots $r_h^-$ and $r_h^+$ for inner and outer horizons, respectively. In addition, from the extreme condition $r_h^-=r_h^+$ established by the system of equations $f(r)=0$ and $f'(r)=0$, the critical charge parameter is obtained, $l_{\text{ex}}=8m/27=0.296m$. 

\begin{figure}[t]
\begin{minipage}{0.9\linewidth}
\center{\includegraphics[width=0.97\linewidth]{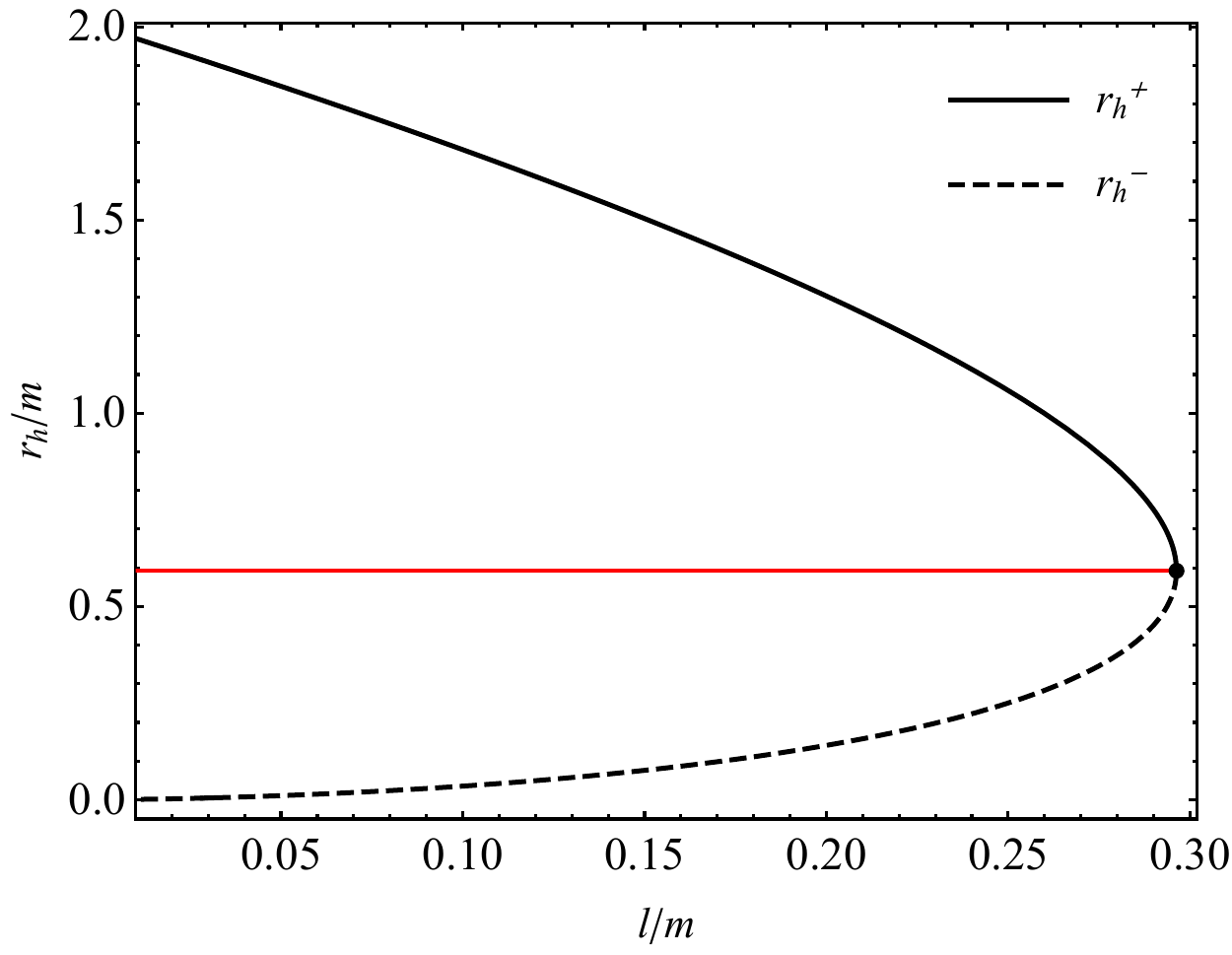}\\ } 
\end{minipage}
\hfill 
\caption{Dependence of normalized horizons $r_h^-/m$ and $r_h^+/m$ on the charge parameter $l/m$. The boundary between inner and outer horizons is marked by red line. The black dot indicates the extreme BH case, namely, $l/m=8/27$, $r_h/m=0.59$.}
\label{fig:horizons}
\end{figure}

The numerical behavior of the horizons is shown in Fig.~\ref{fig:horizons}. As the charge parameter increases, the inner and outer horizons approach each other, eventually merging at $l_{\text{ex}}/m$. The case $l>l_{\text{ex}}$ geometrically describes a compact object without an event horizon, implying the absence of a BH in the traditional sense.

A comprehensive analysis of the Fan-Wang RBH horizons, along with the detailed behavior of the lapse function, can be found in \cite{2024PDU....4601566K}.

\subsection{Curvature invariants}

When analyzing the geometric properties of the Fan-Wang RBH, curvature invariants help to understand the spacetime structure, especially, for identifying regions of strong gravitational fields and for indicating the regularity of spacetime. The Ricci scalar $R$ serves as a measure of curvature, which is directly influenced by the local distribution of matter and energy. The square of the Ricci tensor $R_{\alpha\beta}R^{\alpha\beta}$ provides additional details, including the intensity of the curvature throughout the entire Ricci tensor structure. The Kretschmann invariant $K$, on the other hand, reflects the total curvature of spacetime (both in the vicinity of a BH and inside it). This invariant is particularly useful for detecting singularities, as it often diverges in classical BH solutions at $r=0$. For the Fan-Wang spacetime, specified invariants read:
\begin{equation}\label{ricci}
    R=g^{\alpha\beta}R_{\alpha\beta}=\frac{24l^2m}{(l+r)^5},
\end{equation}
\begin{equation}\label{ricci_sq}
    R_{\alpha\beta}R^{\alpha\beta}= \frac{144l^2m^2(l^2+r^2)}{(l+r)^{10}},
\end{equation}
\begin{equation}\label{kretsch}
K=R_{\alpha\beta\mu\nu}R^{\alpha\beta\mu\nu}=\frac{48m^2(2l^4+7l^2r^2-2lr^3+r^4)}{(l+r)^{10}},
\end{equation}
where $R_{\alpha\beta\mu\nu}$ is the Riemann tensor. In the case of $l=0$, the Ricci tensor vanishes by definition, while the Kretschmann scalar reduces to a value $K=48m^2/r^6$ corresponding to the Schwarzschild spacetime. At the point $r=0$, curvature invariants are finite, namely,
\begin{eqnarray}
    \lim\limits_{r \to 0} R &=&\frac{24m}{l^3}, \quad \lim\limits_{r \to 0} R_{\alpha\beta}R^{\alpha\beta} =\frac{144m^2}{l^6}, \nonumber\\  \lim\limits_{r \to 0} K &=& \frac{96m^2}{l^6}.
\end{eqnarray}

\begin{figure*}[t]
\begin{minipage}{0.505\linewidth}
\center{\includegraphics[width=0.97\linewidth]{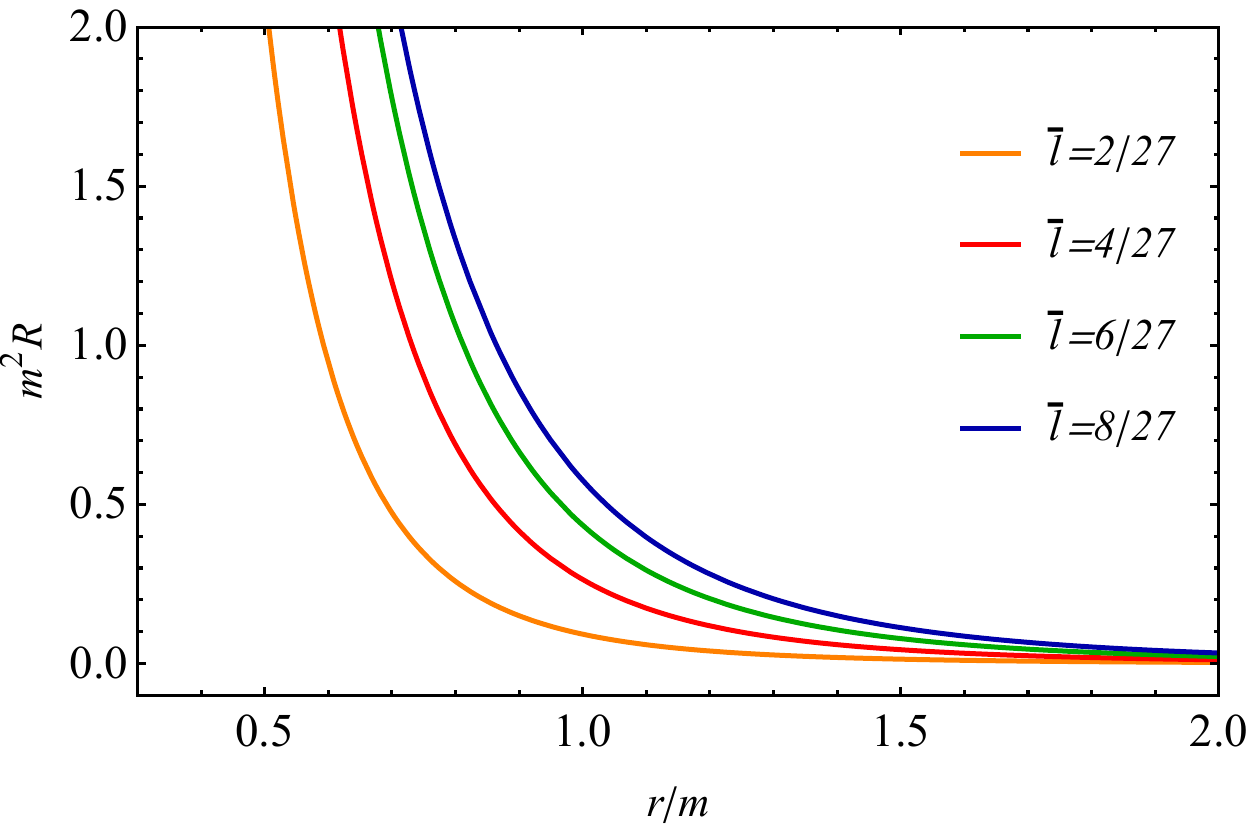}\\ } 
\end{minipage}
\hfill 
\begin{minipage}{0.485\linewidth}
\center{\includegraphics[width=0.97\linewidth]{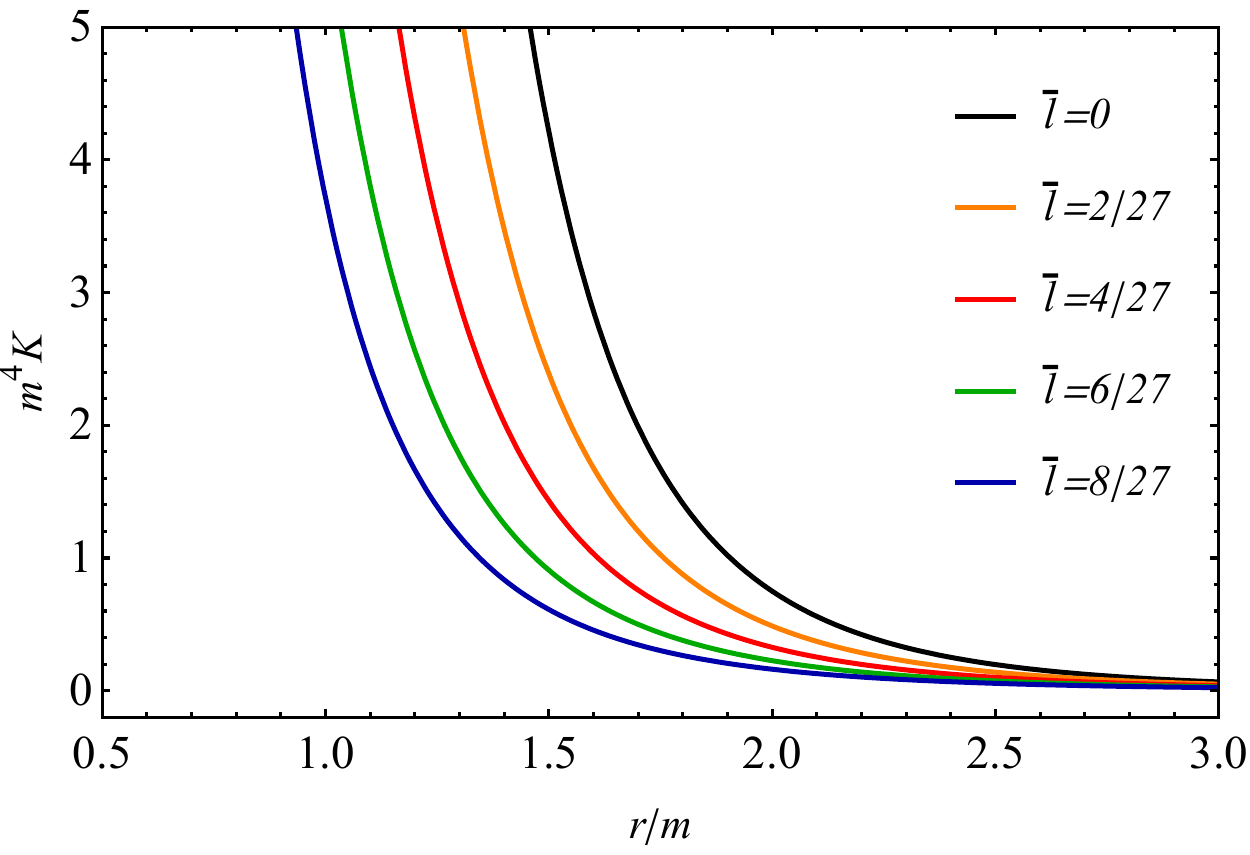}\\ }
\end{minipage}
\caption{Dimensionless curvature invariants as functions of dimensionless radial coordinate $r/m$ for different values of charge parameter $\bar{l}=l/m$. Left panel: Ricci scalar, Middle panel: Ricci tensor squared, Right panel: Kretschmann scalar.}
\label{fig:curv_invars}
\end{figure*}

Fig.~\ref{fig:curv_invars} illustrates the radial profiles of dimensionless curvature invariants for various values of the charge parameter. The distributions of the Ricci scalar $R$ and the square of the Ricci tensor $R_{\alpha\beta}$ exhibit similar trends, with increasing $\bar{l}=l/m$ leading to higher values of these invariants. Conversely, the Kretschmann scalar decreases as $\bar{l}$ increases, highlighting the distinct manner in which the total curvature responds to changes in the BH's charge parameter.

\subsection{Energy conditions}
A detailed discussion of the energy conditions for the Fan-Wang RBH spacetime is presented in Ref.~\cite{2022JHEP...11..108M}, where the physical properties of the energy-momentum tensor $T_{\mu}^{\nu}=\text{diag}(-\rho,P_1,P_2,P_3)$ are examined. The Fan-Wang RBH satisfies standard energy conditions, including the null energy condition (NEC), weak energy condition (WEC), and dominant energy condition (DEC).
The NEC, $\rho+P_i \geq 0$, is satisfied throughout spacetime, ensuring that light-like geodesics behave physically meaningfully. The WEC, $\rho \geq 0$, which implies that the energy density as measured by any observer is non-negative, is also satisfied everywhere. Similarly, the DEC, $\rho-P_i \geq 0$, guarantees that the energy flux never exceeds the speed of light, thus preserving causality.

One of the key observations relates to the strong energy condition (SEC), $\rho+\sum_{i=1}^{3}P_i\geq 0$, which is typically used to describe the attractive nature of gravity and its tendency to focus on geodesics. For the Fan-Wang spacetime, the SEC is defined by:
\begin{equation}\label{SEC}
   \rho+P_1+2P_2 = \frac{12ml(r-l)}{(r+l)^5},
\end{equation}
where $\rho$ is the energy density, $P_1=P_r$ and $P_2=P_{\theta}=P_3=P_{\phi}$ are the radial and angular pressures. This expression reveals that the SEC is violated in the region $0\leq r < l$, as the right-hand side becomes negative when $r < l$. The violation of the SEC in this region indicates that the gravitational field becomes repulsive and prevents the formation of a singularity.

\section{Geodesic motion}\label{sec:geod_motion}

The motion of test particles is of utmost importance as it may represent a direct signature of the gravitational field induced by a given spacetime. For example, the geodesic motion is associated with experimental evidence such as quasiperiodic oscillations \cite{2023PhRvD.108d4063B,2023PhRvD.108l4034B} and, therefore, it appears crucial to find out the main deviations of the Fan-Wang solution from the standard Schwarzschild spacetime. Here, we focus on massive and non-massive particles, including photons.

\subsection{Dynamics of massive particles} \label{sec:dyn_massive}

The equations of motion are derived from the Lagrangian of a test particle \cite{shapiro2008black}, which in the case of motion on the equatorial plane ($\theta=\pi/2, \dot{\theta}=0$) takes the form\footnote{In this work, we restrict ourselves to the motion on the equatorial plane.}:
\begin{eqnarray} \label{lagrangian}
    \mathcal{L}&=&\frac{1}{2}g_{\alpha\beta}\dot{x}^{\alpha}\dot{x}^{\beta}=\frac{1}{2}\Big[-\left(1-\frac{2M(r)}{r}\right)\dot{t}^2 \nonumber\\ 
    &+&\left(1-\frac{2M(r)}{r}\right)^{-1}\dot{r}^2 + r^2\dot{\phi}^2\Big],
\end{eqnarray}
where $\dot{x}^{\alpha}=u^{\alpha} = dx^{\alpha}/d\tau$ ($\alpha, \beta = 0,1,2,3$) is 
the 4-velocity vector, and $\tau$ is the proper time for massive particles moving along  timelike geodesics and affine parameter in case of null  geodesics, respectively. Since the Lagrangian \eqref{lagrangian} does not depend on the cyclic coordinates ($t,\phi$),  the corresponding conjugate momenta ($p_t,p_{\phi}$) are conserved. Then, the Euler-Lagrange equations
\begin{equation} \label{EL_eqs}
    \frac{d}{d\tau}\left(\frac{\partial \mathcal{L}}{\partial \dot{x}^{\alpha}}\right) - \frac{\partial \mathcal{L}}{\partial x^{\alpha}} = 0
\end{equation}
and the normalization condition
\begin{equation} \label{normalization}
    g_{\alpha \beta}p^{\alpha} p^{\beta} = -k\mu^2,
\end{equation}
where $\mu$ is a particle mass, yield the corresponding equations of motion. For timelike geodesics ($k=1$),  one obtains
\begin{eqnarray}
\label{eq_t}
   p_{t}&=&\frac{\partial \mathcal{L}}{\partial \dot{t}}=-E=-\left(1-\frac{2M(r)}{r}\right)\dot{t},\\
\label{eq_phi}
   p_{\phi}&=&\frac{\partial \mathcal{L}}{\partial \dot{\phi}}=L = r^2\dot{\phi},\\
   \label{eq_r}
   \dot{r}^2&=&E^2 - \left(1-\frac{2M(r)}{r}\right)\left(1+\frac{L^2}{r^2}\right).
\end{eqnarray}
Here, $E$ and $L$ are integration constants associated with the total energy and angular momentum of a neutral test particle, respectively. Eq.~\eqref{eq_r} can be presented in terms of the effective potential $V_{eff}$
\begin{equation}
    \dot{r}^2 + V_{eff} = \frac{E^2}{\mu^2},
\end{equation}
which, particularly, for massive particles is given by
\begin{equation}
    V_{eff} =  \left(1-\frac{2M(r)}{r}\right)\left(1+\frac{\tilde{L}^2}{r^2}\right),
\end{equation}
here $\tilde{L}=L/\mu$ is a specific angular momentum. Analogously, the specific energy $\tilde{E}=E/\mu$ is introduced below.
The conditions
\begin{equation} \label{circ_orb_cond}
    \dot{r}=0,\quad \quad \frac{dV_{eff}}{dr}=0.
\end{equation}
define $\tilde{E}$ and $\tilde{L}$ of a particle on a circular geodesic orbit around the Fan--Wang BH at a radius $r$, namely:
\begin{eqnarray}
\label{ang_mom_circ}
   \tilde{L}^2(r)&=&\frac{mr^4(r-2l)}{(l+r)^4-3mr^3}, \\
   \label{energy_circ}
   \tilde{E}^2(r)&=& \left(1-\frac{2mr^2}{(l+r)^3}\right)\left[1+\frac{mr^2(r-2l)}{(l+r)^4-3mr^3}\right].
\end{eqnarray}

A detailed analysis of the circular motion around the Fan--Wang BH can be found in \cite{2023PhRvD.107h4003I} and \cite{2024PDU....4601566K}, where the behavior of the energy $\tilde{E}$, angular momentum $\tilde{L}$, and radii of the ISCO were investigated.

\subsection{Dynamics of massless particles} \label{sec:dyn_massles}

The motion of massless particles, such as ``neutrino-like''\footnote{The term ``neutrino-like'' particle is treated as a reference in analyzing optical phenomena related to photon propagation.} particles that have an approximately zero rest mass-energy ($\mu=0$), is governed by the null geodesics of the NED RBH spacetimes \cite{2019ApJ...887..145S}. It is well known that photons do not propagate along null geodesics of spacetime geometry because of the nonlinearity of the electromagnetic field \cite{PhysRevD.61.045001,2019ApJ...874...12S,2019ApJ...887..145S,2019EPJC...79...44S,2019EPJC...79..988S}. Instead of this, the photon motion is governed by the so-called effective geometry (see Sect. \ref{sec:dyn_photon}).

For massless particles following null geodesics ($k=0$ in \eqref{normalization}), the equations of motion become
\begin{eqnarray}
\label{eq_t_m0}
   \frac{dt}{d\tau}&=&E\left(1-\frac{2M(r)}{r}\right)^{-1},\\
\label{eq_phi_m0}
   \frac{d\phi}{d\tau}&=&\frac{L}{r^2},\\
   \label{eq_r_m0}
   \left(\frac{dr}{d\tau}\right)^2&=&E^2 - \frac{L^2}{r^2}\left(1-\frac{2M(r)}{r}\right).
\end{eqnarray}
Replacing a parameter $\tau \rightarrow L\tau$ transforms Eq.~\eqref{eq_r_m0} into
\begin{equation}\label{eq2_r_m0}
    \left(\frac{dr}{d\tau}\right)^2 = \frac{1}{b^2} - V_{eff},
\end{equation}
where  $b=L/E$ is the impact parameter of particles and the effective potential $V_{eff}$ reads
\begin{equation} \label{Veff_m0}
    V_{eff}=\frac{1}{r^2}\left(1-\frac{2M(r)}{r}\right).
\end{equation}

Circular orbits are determined in a standard way by imposing the conditions $dr/d\tau=0$, $dV_{eff}/dr=0$ or, equivalently, $\tilde{L} = \infty$, leading to an expression for the corresponding radius $r_c$:
\begin{eqnarray}\label{circ_orb_m0}
    r_c^{\pm} &=& \frac{1}{4}\Bigg\{3m-4l + \Big[N + 2(H+G)\Big]^{\frac{1}{2}}\nonumber\\ &\pm& \left[2\left(N + \frac{K}{\sqrt{N + 2(H+G)}} -(H+G)\right)\right]^{\frac{1}{2}} \Bigg\},
\end{eqnarray}
where
\begin{eqnarray}
    \label{rc_H}
   H&=&2l^3m/G, \\
   \label{rc_G}
   G&=& \left[4m^{3/2}l^4\left(9\sqrt{m} - \sqrt{81m-256l}\right)\right]^{1/3},\\
   \label{rc_K}
    K&=& 9m(8l^2-12lm+3m^2),\\
    \label{rc_N}
    N&=& 3m(3m-8l).
\end{eqnarray} 

\begin{figure*}[ht]
\begin{minipage}{0.49\linewidth}
\center{\includegraphics[width=0.95\linewidth]{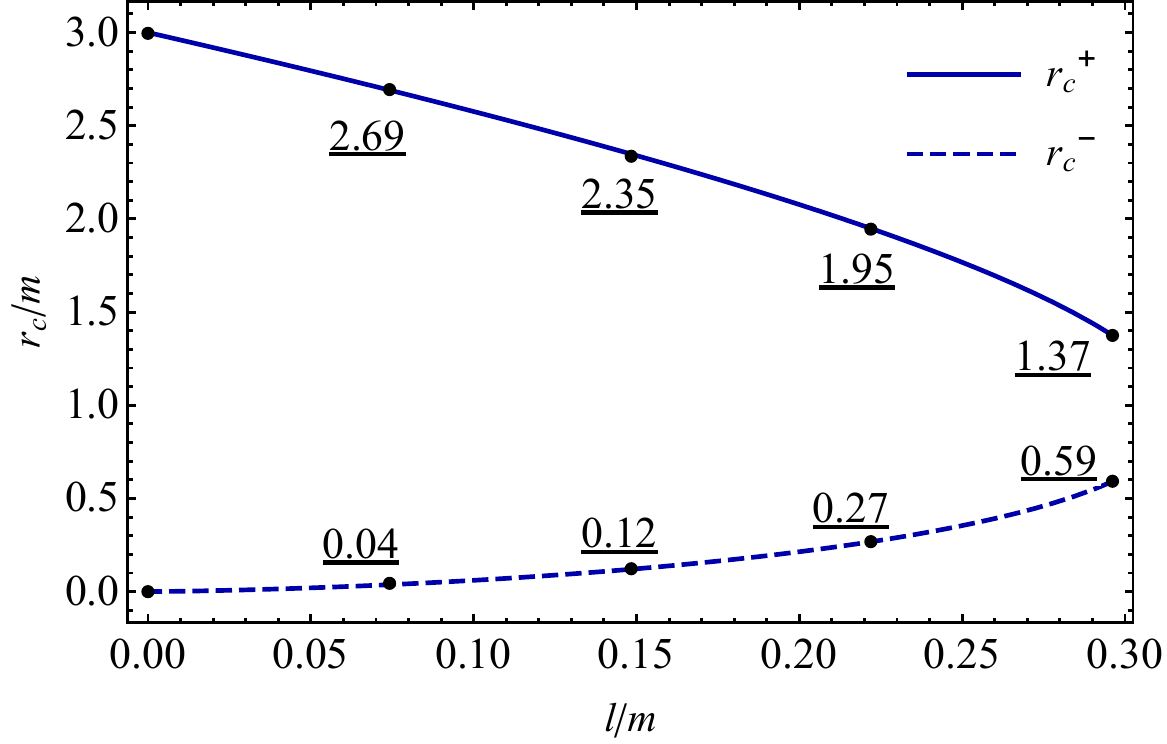}\\ }
\end{minipage}
\hfill 
\begin{minipage}{0.50\linewidth}
\center{\includegraphics[width=0.99\linewidth]{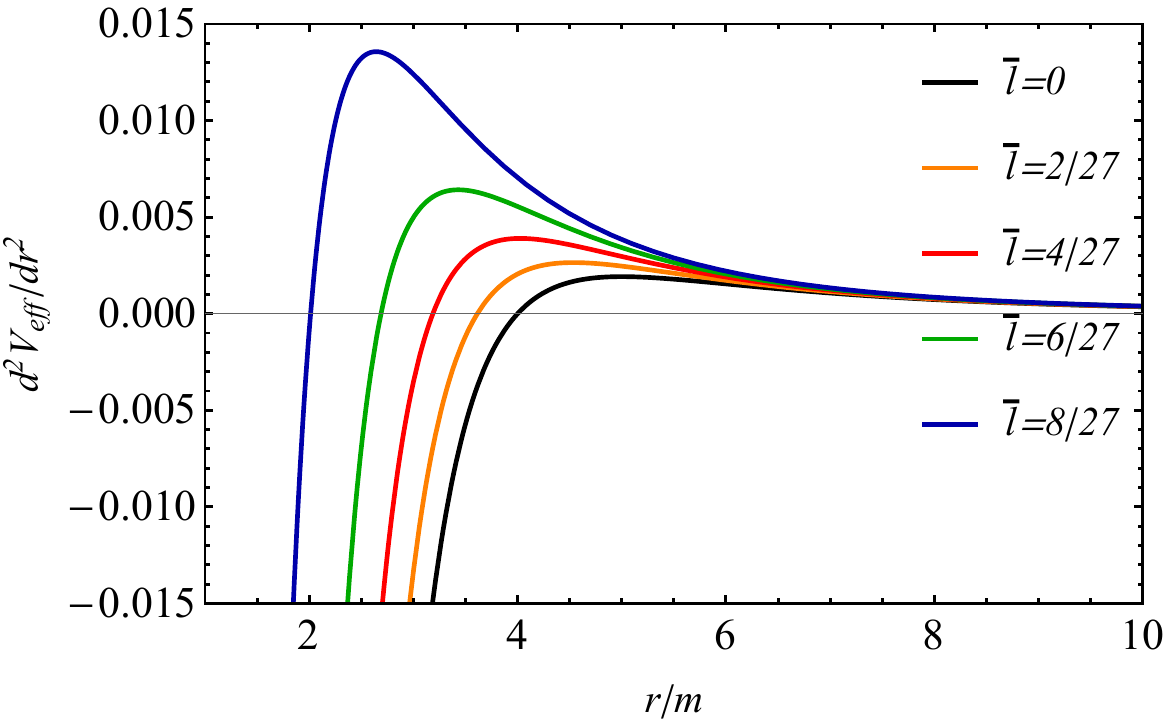}\\ }
\end{minipage}
\caption{Left panel: The normalized radius of the circular orbit $r^{\pm}_c/m$ of massless particles  as a function of the dimensionless charge parameter $l/m$. Dots with underlined numbers indicate the values of $r_c^+/m$ corresponding to $l/m=\{2/27, 4/27, 6/27, 8/27\}$. Right panel: Second derivative of the effective potential $d^2V_{eff}$ for massless particles as a function of normalized radial distance $r/m$ for different values of $\tilde l =l/m$.}
\label{fig:circorb_m0}
\end{figure*}

The left panel of Fig.~\ref{fig:circorb_m0} shows that the radius of the circular orbit ($r_c^+$) decreases with increasing charge parameter.  The values of $r^-_c$ lie below the inner horizon and correspond to unstable circular orbits. In the limiting case $l/m=0$, one obtains the radius of the photon sphere in the Schwarzschild spacetime $r_c/m=3$. 

Moreover, the stability of circular orbits can be estimated through the behavior of the second derivative of the effective potential, i. e., 
\begin{equation}
    \frac{d^2V_{eff}}{dr^2} = \frac{(l+r)^5 - 4mr^4}{r^4 (l+r)^5}.
\end{equation}

The condition $d^2V_{eff}/dr^2>0$ indicates that the orbits are stable, while $d^2V_{eff}/dr^2<0$ implies the opposite. In the right panel of Fig.~\ref{fig:circorb_m0}, the stability regions are shown for different values of the charge parameter.

\subsection{Dynamics of photons} \label{sec:dyn_photon}

 As mentioned above, photons follow  null geodesics of an effective geometry that directly reflects the effects of NED, in addition to those effects embedded within the overall spacetime structure. In this context, the effective metric can be presented as:
\begin{equation}\label{eff_metr_tens}
    \tilde{g}^{\mu\nu} = \mathcal{L}_{\mathcal{F}} g^{\mu\nu}-4\mathcal{L}_{\mathcal{FF}}\mathcal{F}_{\alpha}^{ \mu}\mathcal{F}^{\alpha\nu},
\end{equation}
where $\mathcal{L}_{\mathcal{F}}=\partial \mathcal{L}/\partial\mathcal{F}$, $\mathcal{L}_{\mathcal{FF}} = \partial^2\mathcal{L}/\partial\mathcal{F}^2$. The corresponding line element takes the form \cite{2023PhRvD.107h4003I}:
\begin{equation}\label{eff_metr}
    d\tilde{s}^2 = -\frac{f(r)}{\mathcal{L}_{\mathcal{F}}}dt^2+\frac{1}{ \mathcal{L}_{\mathcal{F}}f(r)}dr^2+\frac{r^2}{\Phi}\left(d\theta^2+\sin^2{\theta}d\phi^2\right),
\end{equation}
and
\begin{equation}
    \Phi=\mathcal{L}_{\mathcal{F}} + 2\mathcal{L}_{\mathcal{F}\mathcal{F}}.
\end{equation}

The effective geometry remains static and spherically symmetric and, therefore, admits the constants of motion related to energy $k_t=-\tilde{E}$ and angular momentum $k_{\phi}=\tilde{L}$, where $k_{\mu}$ is the 4-wave vector associated with the 4-momentum of the photon ($\tilde{p}_{\mu}=\hbar k_{\mu}$). Notice that the constants $\tilde{E}$ and $\tilde{L}$ used here are different from the corresponding constants of motion for massive particles. Employing the eikonal equation for photons
\begin{equation}\label{eikonal_eq}
    \tilde{g}_{\mu\nu}k^{\mu}k^{\nu} = 0,
\end{equation}
the equations of motion can be written as:
\begin{eqnarray}
\label{eq_t_phot}
    \frac{dt}{d\tau} &=& \tilde{E} {\mathcal{L}_\mathcal{F}} \left(1-\frac{2M(r)}{r}\right)^{-1}, \\
    \label{eq_phi_phot}
    \frac{d\phi}{d\tau} &=& \frac{\tilde{L}}{r^2}\Phi, \\
    \label{eq_r_phot}
    \left(\frac{dr}{d\tau}\right)^2 &=& \mathcal{L}_\mathcal{F}^2\left[\frac{1}{\tilde{b}^2} - \frac{\Phi}{r^2 \mathcal{L}_\mathcal{F}}\left(1-\frac{2M(r)}{r}\right)\right],
\end{eqnarray}
where $\tilde{b} = \tilde{L}/\tilde{E}$ is the impact parameter. 
The effective potential $\tilde{V}_{eff}$ is defined in the following way:
\begin{eqnarray}\label{Veff_phot}
    V_{eff} = \frac{\Phi}{r^2}\left(1-\frac{2M(r)}{r}\right).
\end{eqnarray}

\begin{figure*}[ht]
\begin{minipage}{0.49\linewidth}
\center{\includegraphics[width=0.96\linewidth]{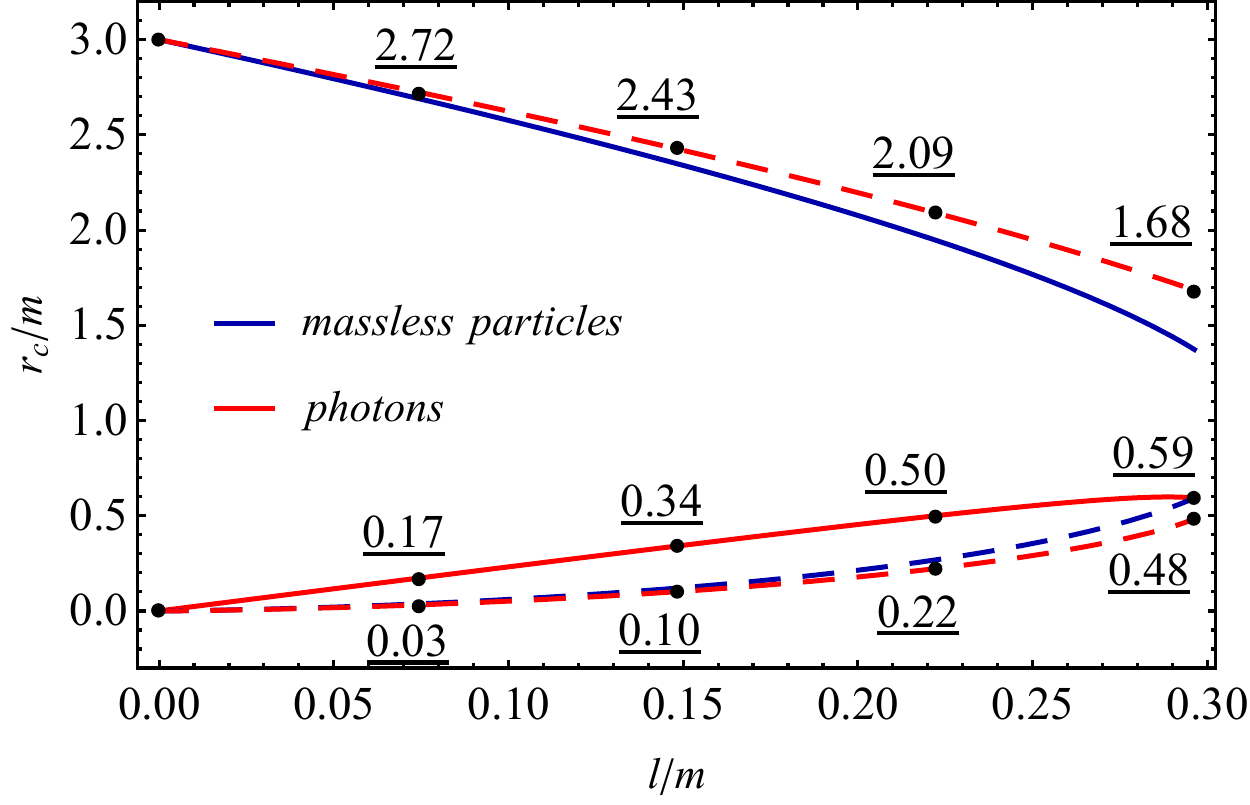}\\ }
\end{minipage}
\hfill 
\begin{minipage}{0.50\linewidth}
\center{\includegraphics[width=0.98\linewidth]{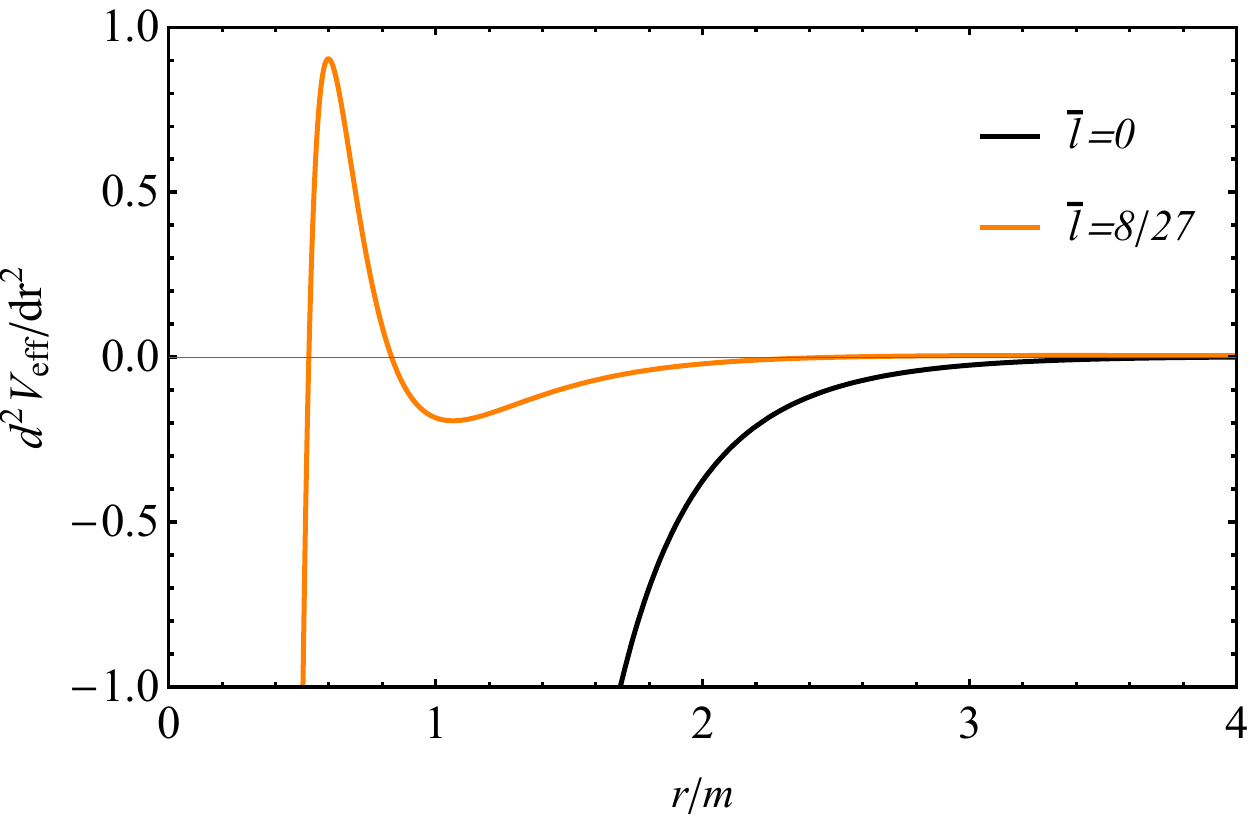}\\ }
\end{minipage}
\caption{Left panel: The normalized radii of circular orbits $r_c/m$ of photons (red) in comparison to those of massless particles (blue)  as functions of dimensionless charge parameter $l/m$. Dashed blue and dashed red lines correspond to unstable orbits. Solid and dashed blue lines are those of Fig.~\ref{fig:circorb_m0}, left. Dots with underlined numbers indicate the values of $r_c$ corresponding to $l/m=\{2/27, 4/27, 6/27, 8/27\}$. Right panel: Second derivative of effective potential $d^2V_{eff}/dr^2$ for photons as a function of normalized radial distance $r/m$ (a representative example with $l/m=8/27$).}
\label{fig:circ_orb_phot}
\end{figure*}

The circular orbits of photons are obtained, as usual, by setting $dr/d\tau=0$ and $dV_{eff}/dr=0$. The equation $dV_{eff}/r=0$ has three real roots. These roots correspond to different branches of circular photon orbits, as illustrated in the left panel of Fig.~\ref{fig:circ_orb_phot}. The upper line represents the photon sphere outside the event horizon, while the middle and lower lines indicate the circular photon orbits located inside the inner horizon. The appearance of these three branches can be attributed to the modified spacetime structure leading to a more intricate effective potential compared to the Schwarschild spacetime. One can check that the effective potential exhibits multiple extrema, with the middle branch corresponding to a local minimum (a stable orbit), while the outer and inner branches correspond to local maxima (unstable orbits).
In the present analysis, we confirm the previous result from \cite{2023PhRvD.107h4003I}. 

Computing the second derivative of the effective potential, we obtain:
\begin{eqnarray}\label{sec_deriv_phot}
    \frac{d^2V_{eff}}{dr^2} &=& \frac{1}{{r^4(l+r)^6}}\Big((l+r)^3(9l^3 + 22l^2r+12lr^2-6r^3)\nonumber\\ &+&4mr^3(6r-19l)\Big).
\end{eqnarray}
Fig.~\ref{fig:circ_orb_phot} (right panel) displays the behavior of  Eq.~\eqref{sec_deriv_phot}. It can be inferred that the outer and inner branches of circular orbits are unstable, while the middle branch is stable.  In addition, photon orbits are compared with orbits of massless particles. The values of $r_c$ in the limiting case $l/m=0$ reduce to the Schwarzschild case for both massless particles and photon orbits (outer ones); nevertheless, the difference between them increases with $l/m$. At the point $r_c/m = 0.59 = r_h^{\pm}$, the unstable orbit of massless particles and the stable photon orbit coincide and cross the horizon $r_h^{\pm}$.

\section{Gravitational Capture} \label{sec:grav_capt}

The gravitational capture is here investigated for the same cases as explored for test particle motion. We summarize the main results and show the main discrepancies with the Schwarzschild solution.

\subsection{Capture of massive particles}\label{sec:capt_massive}

We here discuss the marginally bound circular orbit (MBCO), that is, the critical orbit where a particle is just bound to the BH. The MBCO radius $r_{mb}$ is identified by solving the equation \cite{shapiro2008black}:
\begin{equation} \label{marg_bound_cond}
    V_{eff} = \tilde{E}^2(r_{mb}) =1.
\end{equation}
The only real solution of Eq.~\eqref{marg_bound_cond} for the Fan-Wang BH is given by:
\begin{eqnarray}\label{mb_r}
    r_{mb}&=& \frac{1}{3}\Bigg[2(2m-3l)\nonumber \\ &+& \frac{A}{\left(B + 6\sqrt{3C}\right)^{1/3}} + \left(B + 6\sqrt{3C}\right)^{1/3}\Bigg],
\end{eqnarray}
where
\begin{eqnarray}
    \label{mb_A}
   A&=&9l^2 - 48lm + 16m^2, \\
   \label{mb_B}
   B&=& 64m^3-9l(3l^2 - 30lm + 32m^2)\\
   \label{mb_C}
    C&=& l^3m[9l(23m-3l)-64m^2].
\end{eqnarray}

In this connection, an important quantity is the impact parameter, i.e., the perpendicular distance between the direction of the incoming particle and the center of the BH, which is defined as  $b=\lim\limits_{r \to \infty} r\sin \phi$. It determines the type of orbit that a particle will follow when it approaches the BH. For a given energy, there is a critical impact parameter $b_{cr}$ that corresponds to the MBO. Particles with impact parameters larger than $b_{cr}$ will escape, while those with smaller impact parameters will be captured by the BH. The critical impact parameter is directly related to the capture cross-section $\sigma_{capt}$ for particles falling from infinity through the relation:
\begin{equation} \label{capture}
    \sigma_{capt}=\pi b_{cr}^2.
\end{equation}

\begin{figure*}[ht]
\begin{minipage}{0.49\linewidth}
\center{\includegraphics[width=0.98\linewidth]{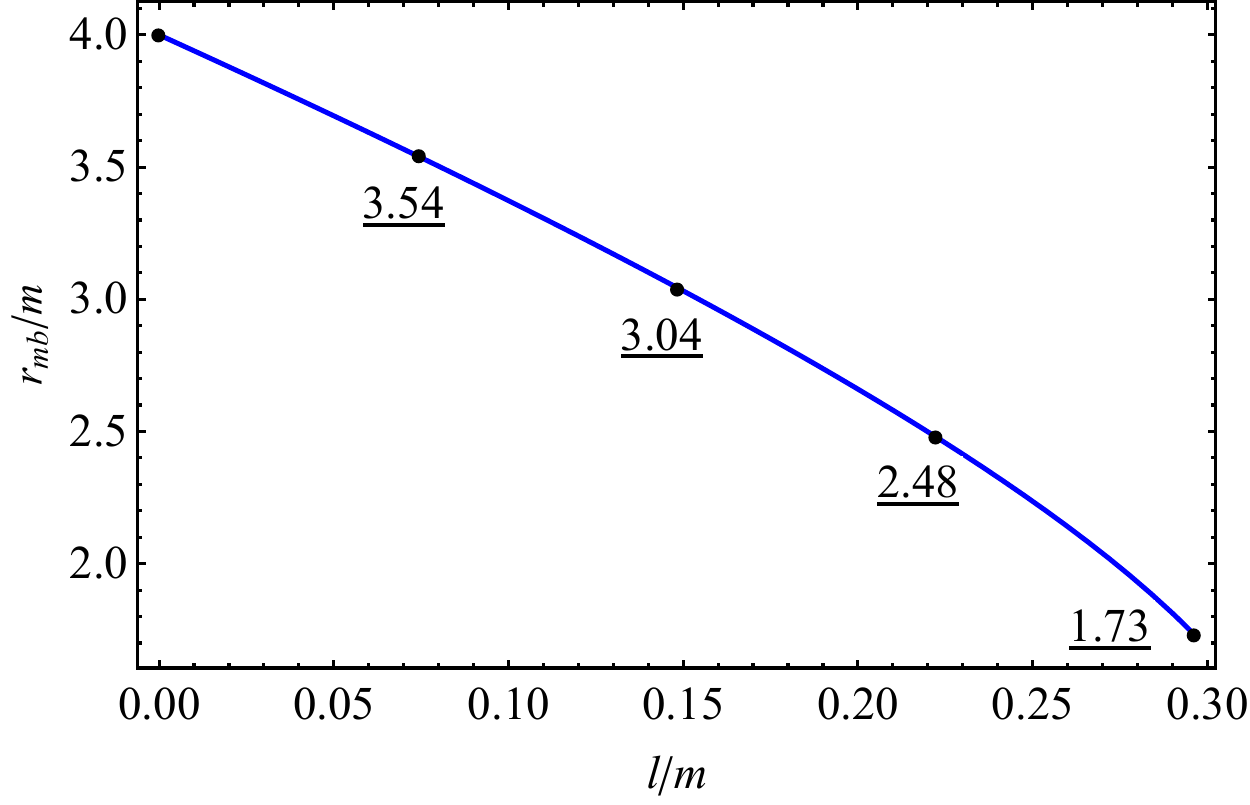}\\ }
\end{minipage}
\hfill 
\begin{minipage}{0.50\linewidth}
\center{\includegraphics[width=0.97\linewidth]{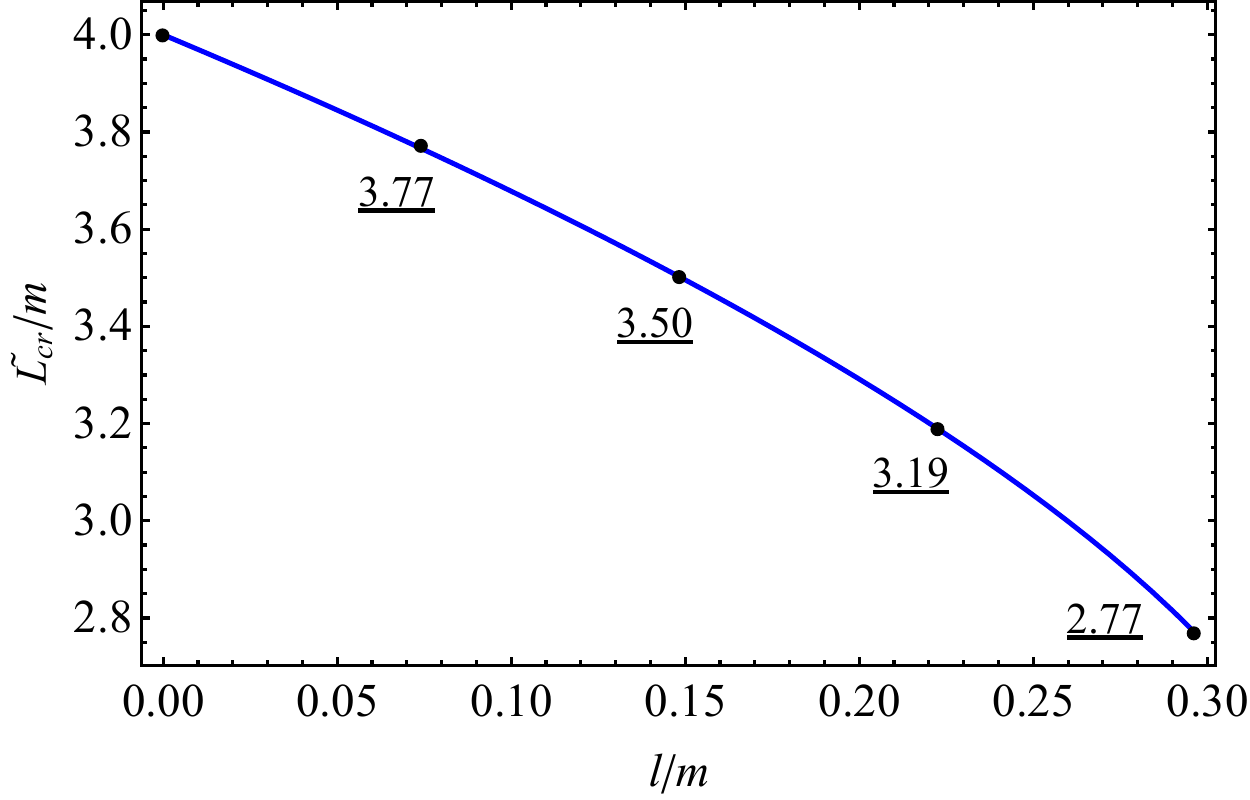}\\ }
\end{minipage}
\caption{Left panel: The normalized radius of marginally bound orbits $r_{mb}/m$ as a function of dimensionless charge parameter $l/m$. Dots with underlined numbers indicate the values of $r_{mb}/m$ corresponding to $l/m=\{2/27, 4/27, 6/27, 8/27\}$. Right panel: The normalized critical angular momentum of a particle $\tilde{L}_{cr}/m$ as a function of dimensionless charge parameter $l/m$. Dots with underlined numbers indicate the values of $\tilde{L}_{cr}$ corresponding to $l/m=\{2/27, 4/27, 6/27, 8/27\}$.}
\label{fig:marg_bound_rL}
\end{figure*}

Assuming $r \to \infty$, Eqs.\eqref{eq_t} and \eqref{eq_phi} together with the definition of $b$ yield \cite{shapiro2008black}
\begin{equation} \label{impact_par}
    b^2_{cr} = \frac{\tilde{L}_{cr}^2}{\tilde{E}^2 - 1},
\end{equation}
where $\tilde{L}_{cr}$ is the critical angular momentum of a particle obtained by substituting Eq.~\eqref{mb_r} in Eq.\eqref{ang_mom_circ}. For Fan--Wang BHs, the behavior of $r_{mb}/m$ and $\tilde{L}_{cr}/m$ as functions of the charge parameter $l/m$ is illustrated in Fig.~\ref{fig:marg_bound_rL}. Note that at $l/m=0$, we have the values $r_{mb}=4m$ and $\tilde{L}_{cr}=4m$, corresponding to the Schwarzschild solution, while for the critical BH case $l/m=8/27$, $r_{mb}=1.73m$ and $\tilde{L}_{cr}=2.77m$.

\begin{figure*}[ht]
\begin{minipage}{0.49\linewidth}
\center{\includegraphics[width=0.95\linewidth]{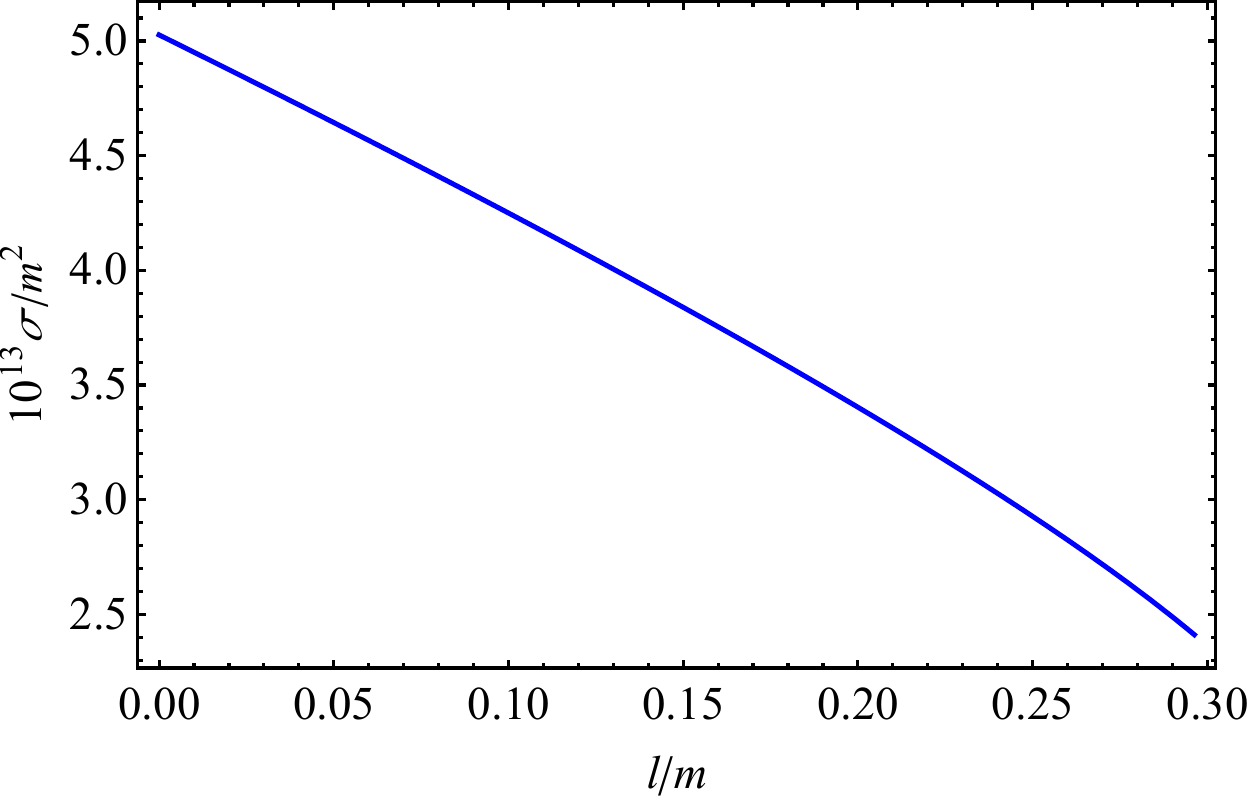}\\ }
\end{minipage}
\hfill 
\begin{minipage}{0.50\linewidth}
\center{\includegraphics[width=0.99\linewidth]{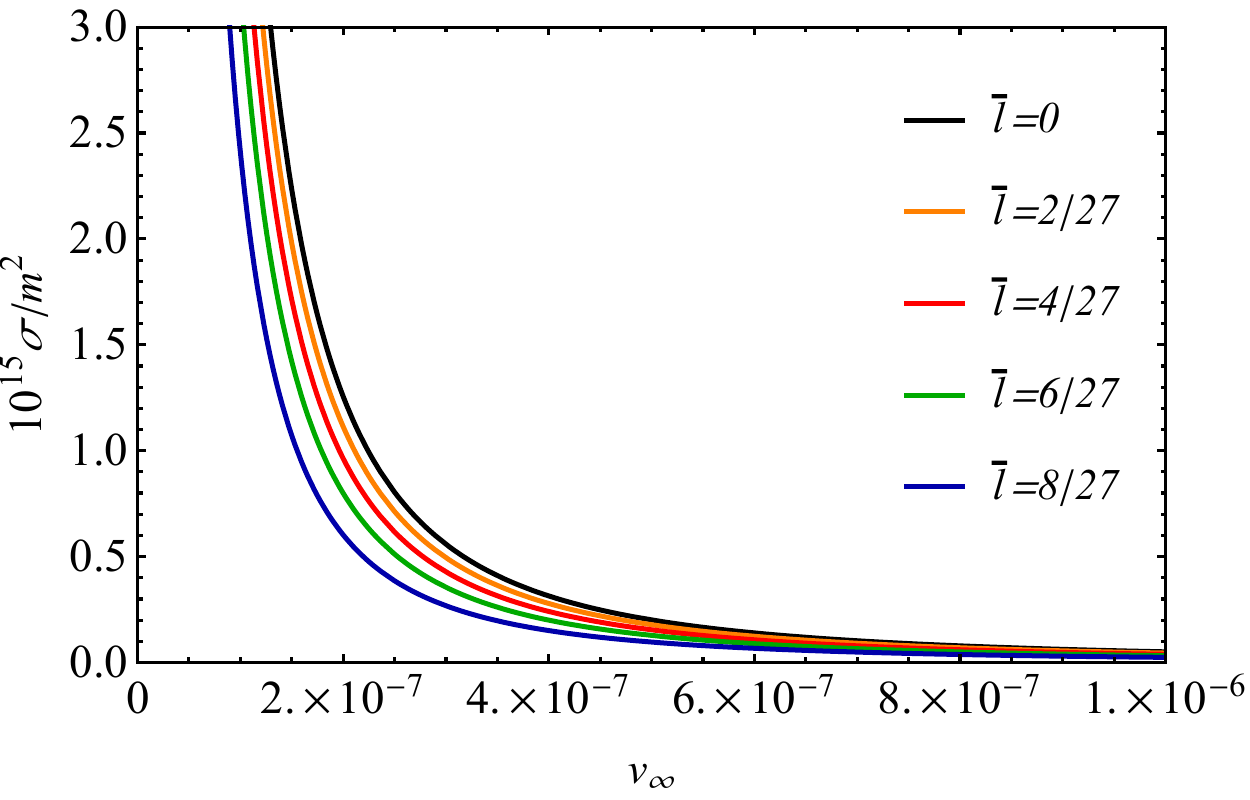}\\ }
\end{minipage}
\caption{Left panel: Dependence of capture cross-section $\sigma/m^2$ on the dimensionless charge parameter $l/m$ (here $v_{\infty}=10^{-6}$ as an example, which corresponds to 0.3 km/s in physical units). Right panel: Dependence of capture cross-section $\sigma/m^2$ on the velocity of non-relativistic particle $v_{\infty}$ (here $\bar{l}=l/m$).}
\label{fig:capture_FW}
\end{figure*}

Furthermore, in the case of non-relativistic particles, Eq.~\eqref{impact_par} can be rewritten in terms of the velocity at infinity $v_{\infty}$, where $\tilde{E}=\left(1-v_{\infty}^2\right)^{-1/2}$ ($v_{\infty}\ll 1$):
\begin{equation}
    b_{cr} = \frac{\tilde{L}_{cr}}{v_{\infty}}.
\end{equation}
In the left panel of Fig.~\ref{fig:capture_FW}, it is demonstrated that with increasing charge parameter $l/m$ up to its critical value, the probability of capturing a particle decreases. The right panel shows that, for different values of $l/m$, the capture 
cross-section decreases with increasing particle velocity. The analytical expression for the cross-section is complicated, so we present it only graphically.

\subsection{Capture of massless particles and photons}\label{sec:capt_massless}

The critical impact parameter of massless particles $b_{cr}$ is obtained by substituting  Eq.~\eqref{circ_orb_m0} in Eq.~\eqref{eq2_r_m0} and setting $dr/d\tau=0$
\begin{equation}\label{capt_massless}
    b_{cr}^2 = r^2 \left.\left(1-\frac{2M(r)}{r}\right)^{-1}\right|_{r=r_c(\text{massless})},
\end{equation}
allowing to calculate the capture cross-section from Eq.~\eqref{capture}. 

\begin{figure*}[ht]
\begin{minipage}{0.50\linewidth}
\center{\includegraphics[width=0.97\linewidth]{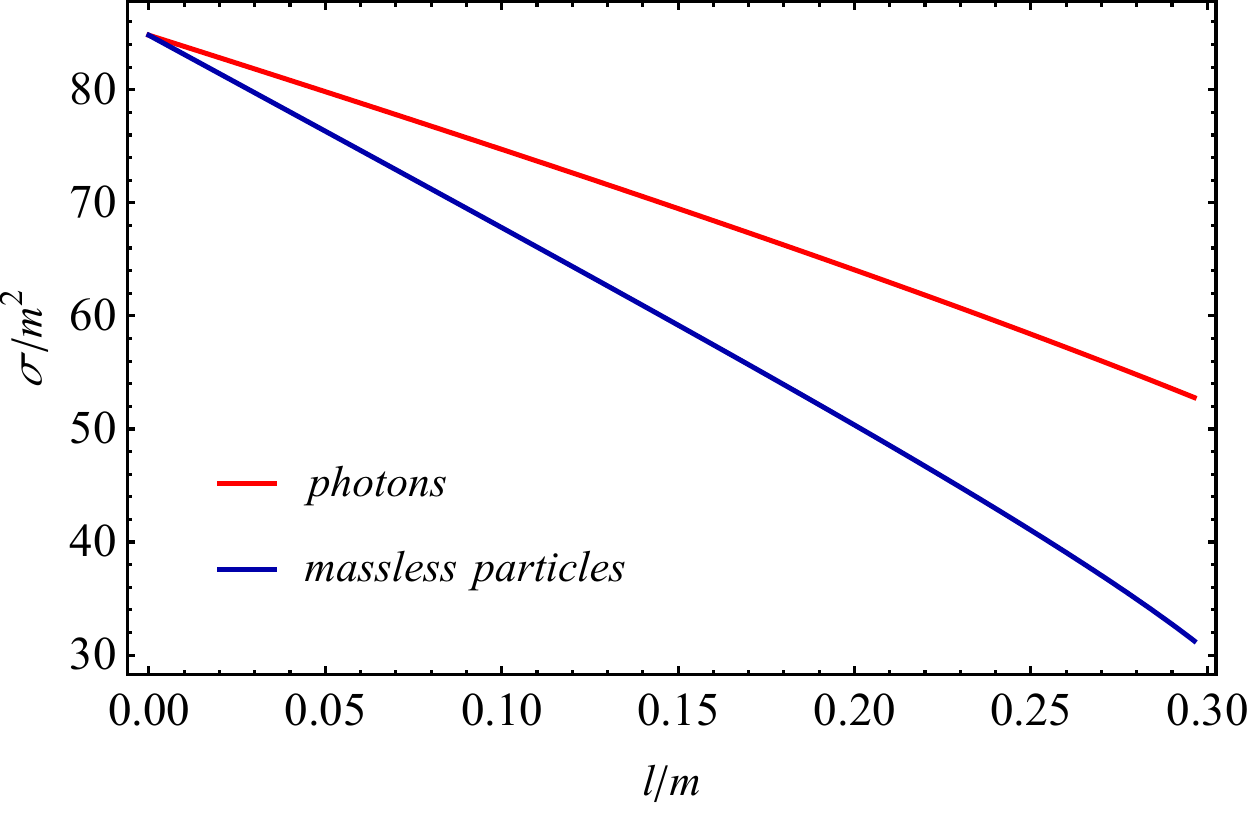}\\ }
\end{minipage}
\hfill 
\begin{minipage}{0.49\linewidth}
\center{\includegraphics[width=0.97\linewidth]{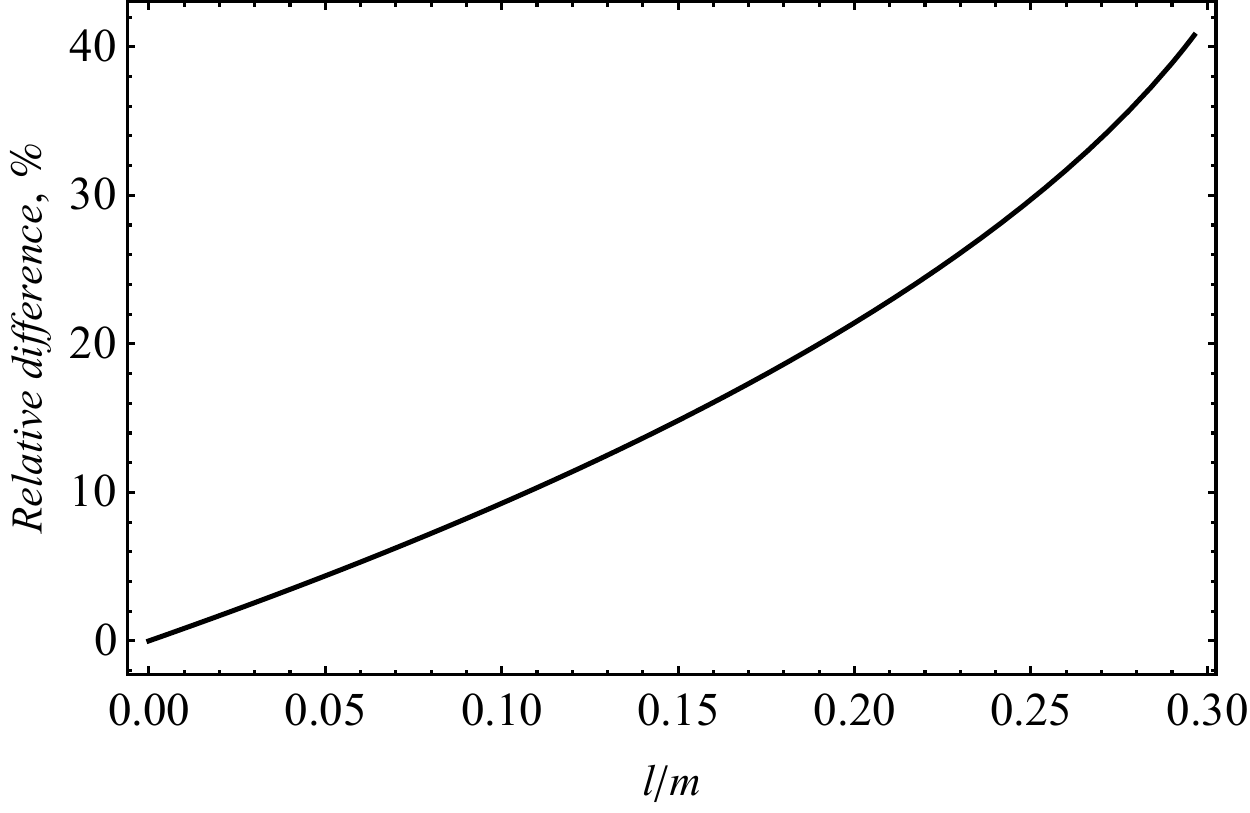}\\ }
\end{minipage}
\caption{Left panel: The normalized capture cross-section $\sigma_{capt}/m^2$ for massless particles and photons as function of dimensionless charge parameter $l/m$. Right panel: Relative difference between capture cross-sections for massless particles and photons.}
\label{fig:capt(m0+phot)}
\end{figure*}

As for photons, setting $dr/d\tau=0$ entails the condition \cite{2019ApJ...874...12S, 2019ApJ...887..145S}:
\begin{equation}
    \mathcal{L}_{\mathcal{F}} - \frac{\Phi \tilde{b}^2}{r^2}\left(1-\frac{2M(r)}{r}\right) = 0,
\end{equation}
from which the critical impact parameter of photons is defined as:
\begin{equation}\label{capt_photons}
    \tilde{b}_{cr}^2 = \frac{\mathcal{L}_{\mathcal{F}} r^2}{\Phi}\left.\left(1-\frac{2M(r)}{r}\right)^{-1}\right|_{r=r_c(\text{photons})}.
\end{equation}
In Eqs.~\eqref{capt_massless} and \eqref{capt_photons}, $r_c$ (massless) and $r_c$ (photons) are the radii of circular orbits computed in Sec.~\ref{sec:dyn_massles} and Sec.~\ref{sec:dyn_photon} (see right panels of Fig.~\ref{fig:circorb_m0} and Fig.~\ref{fig:circ_orb_phot}). The resulting expressions for $\sigma_{capt}$ are rather complicated, so we present them graphically as functions of  $l/m$ in the left panel of Fig.~\ref{fig:capt(m0+phot)}. For both massless particles and photons, an increase in the parameter $l/m$ leads to a decrease in the values of $\sigma_{capt}/m^2$, whereas for $l/m\rightarrow0$ the Schwarzschild limit $\sigma_{capt}/m^2=27\pi=84.823$ is recovered. It is evident that the difference between the cross-sections for massless particles and photons becomes more pronounced as $l/m$ increases. The relative difference $(\sigma_{\text{photon}}-\sigma_{\text{massless}})/\sigma_{\text{photon}}$ is shown in the right panel of Fig.~\ref{fig:capt(m0+phot)}.

\section{Perihelion Shift} \label{sec:perihelion}

In this section, we analyze the perihelion advance in the trajectories of massive particles. Starting with the radial equation of motion, we express $\dot{r}$ as
\begin{equation}\label{r_dot}
    \dot{r} = \frac{dr}{d\tau} = \frac{dr}{d\phi}\frac{d\phi}{d\tau} = \frac{\tilde{L}}{r^2}\frac{d\phi}{d\tau}.
\end{equation}
From Eqs.~\eqref{eq_r} and ~\eqref{eq_phi}, it is possible to derive a first order differential equation for $r(\phi)$:
\begin{equation}\label{eq_r_phi}
    \frac{1}{r^4} \left(\frac{dr}{d\phi}\right)^2 = \frac{1}{\tilde{L}^2}\left[\tilde{E}^2 - \left(1-\frac{2M(r)}{r}\right)\left(1+\frac{\tilde{L}^2}{r^2}\right)\right].
\end{equation}
Direct integration of Eq.~\eqref{eq_r_phi} yields:
\begin{equation}\label{phi_of_r}
    \phi = \pm \int\frac{1}{r^2}\left[\frac{\tilde{E}^2}{\tilde{L}^2} - \frac{1}{\tilde{L}^2}\left(1-\frac{2M(r)}{r}\right)\left(1+\frac{\tilde{L}^2}{r^2}\right)\right]^{-\frac{1}{2}}dr.
\end{equation}
For orbits close to the horizon, the perihelion precession can be obtained by numerical integration of Eq.~\eqref{phi_of_r} (setting the appropriate orbital parameters into the constants of motion). However, for distant orbits where the weak field limit is satisfied, approximate methods might be utilized. Thus, with the use of the common variable change $u=1/r$, Eq.~\eqref{eq_r_phi} can be rewritten in the following form:

\begin{equation}\label{eq_u_phi_FW}
    \left(\frac{du}{d\phi}\right)^2 + u^2 = \frac{\tilde{E}^2-1}{\tilde{L}^2} + \frac{2mu}{(1+lu)^3}\left(\frac{1}{\tilde{L}^2}+u^2\right).
\end{equation}
The derivative of this equation gives the second order differential equation of motion
\begin{equation} \label{eq2_u_phi_FW}
    \frac{d^2u}{d\phi^2} + u = \frac{m}{(1+lu)^4}\left[\frac{1}{\tilde{L}^2}(1-2lu) + 3u^2\right],
\end{equation}
which can be expanded into series up to first order in parameter $l$ to obtain:
\begin{equation} \label{eq2_u_phi_FW_expanded}
    \frac{d^2u}{d\phi^2} + u \approx \frac{m}{\tilde{L}^2}+3mu^2-6ml\left(\frac{u}{\tilde{L}^2}+2u^3\right).
\end{equation}
On the right-hand side of Eq.~\eqref{eq2_u_phi_FW_expanded}, the first term represents Newton's orbital equation, the second term accounts for the contribution from the Schwarzschild spacetime, and the third term ($\sim l$) reflects the influence of NED.

Following the perturbation method described in \cite{narlikar2010introduction, pollock2003mercury}, we solve the Eq.~\eqref{eq2_u_phi_FW_expanded} by assuming a solution of the form
\begin{equation} \label{u_gen}
    u = u_0 + \epsilon u_1 + O(\epsilon^2),
\end{equation}
where $\epsilon = 3m^2/\tilde{L}^2$ is a small dimensionless parameter. By differentiating Eq.~\eqref{u_gen} and replacing  it in Eq.~\eqref{eq2_u_phi_FW_expanded}, we get:
\begin{eqnarray}\label{eq_u_FW_final}
    u_0''&+& u_0 - \frac{m}{\tilde{L}^2} + \epsilon\Big[ u_1''+u_1 - \frac{\tilde{L}^2u_0^2}{m}\nonumber\\ &-& \frac{2lu_0}{m}\left(1+2\tilde{L}^2u_0^2\right)\Big] + O(\epsilon^2) = 0.
\end{eqnarray}

In the first-order approximation, the  equation 
\begin{equation} \label{eq_Newt}
    u_0''+u_0 - \frac{m}{\tilde{L}^2} = 0
\end{equation}
corresponds to Newtonian orbit with the solution
\begin{equation}\label{sol_Newt}
    u_0 = \frac{m}{\tilde{L}^2}\big(1+e\cos \phi\big),
\end{equation}
where $e$ is the orbit eccentricity. Then, we examine the part of Eq.~\eqref{eq_u_FW_final} with a factor $\sim \epsilon$ taking into account \eqref{sol_Newt}:
\begin{eqnarray}\label{eq_u1}
    u_1''+u_1 &=& \frac{m}{\tilde{L}^2}\left(1+\frac{e^2}{2}-\frac{2l}{m}-\frac{4ml}{\tilde{L}^2}-\frac{6e^2ml}{\tilde{L}^2}\right)\nonumber \\
            &+& \frac{2me}{\tilde{L}^2}\left(1-\frac{l}{m}-\frac{6ml}{\tilde{L}^2}-\frac{3e^2ml}{2\tilde{L}^2}\right)\cos \phi\nonumber\\ &+& \frac{me^2}{2\tilde{L}^2}\left(1-\frac{12ml}{\tilde{L}^2}\right)\cos2\phi.
\end{eqnarray}

We search for the general solution of Eq.~\eqref{eq_u1} in the form 
\begin{equation}
    u_1=A + B\phi \sin \phi + C \cos 2\phi,
\end{equation}
which yields
\begin{eqnarray}
\label{A_coef}
   A=\frac{m}{\tilde{L}^2}\left[1+\frac{e^2}{2} - \frac{2l}{m} - \frac{4ml}{\tilde{L}^2}\left(1+\frac{3e^2}{2}\right)\right],\\
\label{B_coef}
   B= \frac{me}{\tilde{L}^2}\left[1-\frac{l}{m}\left(1+\frac{3m^2}{2\tilde{L}^2}(4+e^2)\right)\right],\\
   \label{C_coef}
   C = -\frac{me^2}{6\tilde{L}^2}\left(1-\frac{12ml}{\tilde{L}^2}\right).
\end{eqnarray}

\begin{figure*}[ht]
\begin{minipage}{0.49\linewidth}
\center{\includegraphics[width=0.98\linewidth]{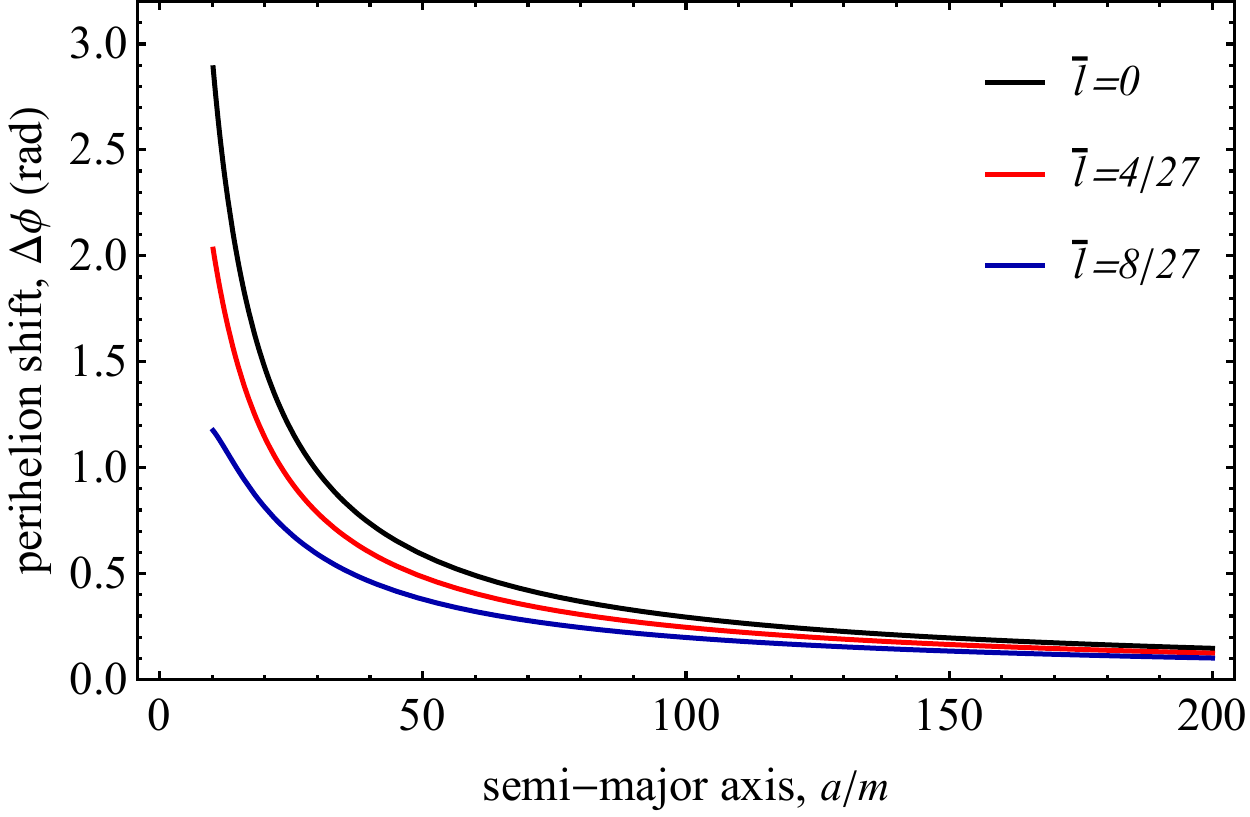}\\ }
\end{minipage}
\hfill 
\begin{minipage}{0.50\linewidth}
\center{\includegraphics[width=0.98\linewidth]{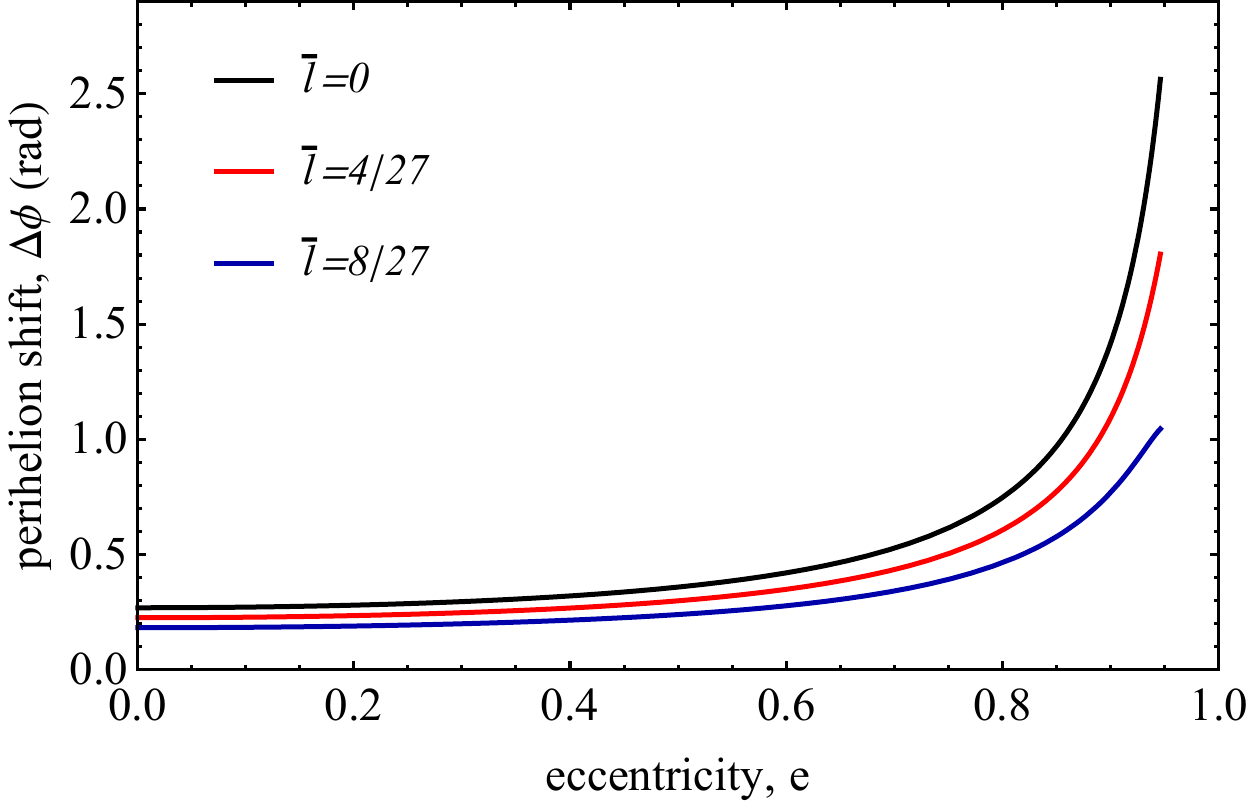}\\ }
\end{minipage}
\caption{Left panel: Perihelion shift $\Delta\phi$ as a function of dimensionless semi-major axis $a$ at a fixed value of orbit eccentricity $e=0.6$ and different values of dimensionless charge parameter $\bar{l}=l/m$. Right panel: Perihelion shift $\Delta\phi$ as a function of eccentricity $e$ at a fixed semi-major axis $a/m=70$ and different values of dimensionless charge parameter $\bar{l}=l/m$.}
\label{fig:perihelion}
\end{figure*}

Afterwards, we write the solution for $u=u_0+\epsilon u_1$ as:
\begin{eqnarray}\label{u_sol}
    u &=& \frac{m}{\tilde{L}^2}\big(1+e\cos \phi\big)\nonumber\\ &+& \epsilon \frac{m}{\tilde{L}^2}\Bigg\{1+\frac{e^2}{2} - \frac{2l}{m} - \frac{4ml}{\tilde{L}^2}\left(1+\frac{3e^2}{2}\right)\nonumber\\ &+&  \left[1-\frac{l}{m}\left(1+\frac{3m^2}{2\tilde{L}^2}(4+e^2)\right) \right]e\phi\sin \phi\nonumber\\ &-& \frac{e^2}{6}\left(1+\frac{2ml}{\tilde{L}^2}\right)\cos 2\phi\Bigg\}.
\end{eqnarray} 
In Eq.~\eqref{u_sol}, the term $\left[1-\frac{l}{m}\left(1+\frac{3m^2}{2\tilde{L}^2}(4+e^2)\right) \right]e\phi\sin \phi$ increases after each revolution and, therefore, becomes dominant. Then, the solution can be written in simplified form as
\begin{eqnarray}
        u &\approx& \frac{m}{\tilde{L}^2}\Bigg[1+e\cos \phi \nonumber\\ &+& \epsilon \left[1-\frac{l}{m}\left(1+\frac{3m^2}{2\tilde{L}^2}(4+e^2)\right) \right]e\phi \sin \phi\Bigg].
\end{eqnarray}
Taking into account that $\epsilon\left[1-\frac{l}{m}\left(1+\frac{3m^2}{2\tilde{L}^2}(4+e^2)\right) \right] \equiv \tilde{\epsilon}$ is a small quantity and using the formula $\cos(\alpha -\beta) = \cos \alpha \cos \beta + \sin \alpha \sin \beta$, we get:
\begin{equation} \label{u_sol_fin}
    u = \frac{m}{\tilde{L}^2}\left\{1+e\cos \left[\phi\left(1-\tilde{\epsilon}\right)\right]\right\}.
\end{equation}
After matching Eq.~\eqref{u_sol_fin} with the solution for the Newtonian orbit in Eq.~\eqref{sol_Newt}, it can be seen that the argument of the cosine modifies by a factor $\delta \phi = \tilde{\epsilon}$. When an orbit completes one full cycle ($2\pi$), the perihelion shift is:
\begin{equation}\label{perih1}
    \Delta \phi = 2\pi(1-\delta \phi) = \frac{6\pi m^2}{\tilde{L}^2} \left[1-\frac{l}{m}\left(1+\frac{3m^2}{2\tilde{L}^2}(4+e^2)\right) \right].
\end{equation}
For an ellipse with semi-major axis $a$, Eq.~\eqref{perih1} can be expressed as
\begin{equation}\label{perih2}
    \Delta \phi = \frac{6\pi m}{a(1-e^2)} \left[1-\frac{l}{m}\left(1+\frac{3m}{2a(1-e^2)} (4+e^2)\right)\right],
\end{equation}
where we used $\tilde{L} = \sqrt{ma(1-e^2)}$. Compared to the classical Schwarzschild result for the perihelion shift $\Delta \phi = 6\pi m/a(1-e^2)$, additional terms arise from the influence of NED effects. The corresponding effects are qualitatively represented in Fig.~\ref{fig:perihelion}, which shows that, for a given $a$ (left panel) and for a given $e$ (right panel), increasing $l$ reduces the magnitude of the perihelion shift. Differences in $\Delta \phi$ are more pronounced for orbits closer to the source and for orbits with higher eccentricity.
In physical units ($m\to Gm/c^2$, $l \to l$), Eq.\eqref{perih2} becomes:
\begin{equation}\label{perih_phys}
    \Delta \phi = \frac{6\pi G m}{c^2 a(1-e^2)} \left[1-\frac{c^2 l}{G m}\left(1 + \frac{3Gm}{2c^2a(1-e^2)}(4+e^2) \right)\right].
\end{equation}

We now apply Eq.~\eqref{perih_phys} to compute the perihelion shift for Mercury as a representative example, to evaluate the influence of the magnetic charge. For the values $l=\{0, 2/27, 4/27, 6/27, 8/27\}$ (in units of length), we obtain $\Delta\phi= \{43.0541, 43.0519, 43.0497, 43.0476, 43.0454\}$ arcseconds per century with the pure correction $\sim l$ having values $\{0,-0.0022, -0.0043, -0.0065, -0.0085\}$ arcseconds per century. Although in the case of Mercury the NED contribution leads to a reduction in the perihelion shift, all computed values remain at the threshold of observational measurements.

\section{Gravitational deflection} \label{sec:grav_defl}

In this section, we analyze the effects induced by the Fan-Wang metric on the gravitational deflection of massless particles and photons. 
 Our purpose is to find out whether the motion of massless particles, such as neutrino-like fields\footnote{Neutrinos are  massive particles, as experimentally found in \cite{2020EPJC...80..964B}. However, for several infrared tests, it is possible to handle them approximately as massless particles.}, can be used to experimentally test the discrepancies induced by this precise regular spacetime.

\subsection{Deflection of massless particles}\label{sec:defl_massless}

Eqs.~\eqref{eq_phi_m0}-\eqref{eq_r_m0} for massless particle motion lead to
\begin{equation}\label{eq_orbit_m0}
    \frac{1}{r^4}\left(\frac{dr}{d\phi}\right)^2 = \left[\frac{1}{b^2} - \frac{1}{r^2}\left(1-\frac{2M(r)}{r}\right)\right].
\end{equation}
From the assumption that massless particles approaching from infinity reach the turning point $R_0$ (the distance of closest approach, where $dr/d\phi=0$) and then return to infinity, the angle $\phi$ is given by
\begin{equation}\label{delta_phi_deflection}
    \phi = \pm \int_{R_0}^\infty \frac{1}{r^2}\left[\frac{1}{b^2} - \frac{1}{r^2}\left(1-\frac{2M(r)}{r}\right)\right]^{-1/2}dr,
\end{equation}
where $1/b^2=f(R_0)/R_0^2$ \cite{PhysRevD.72.104006}.
The total deflection angle $ \hat{\alpha}(R_0)$ is obtained considering the fact that in the absence of a gravitational field, the trajectory of a massless particle is a straight line with $ \phi=\pi$ \cite{1972gcpa.book.....W}. Then,  
\begin{equation}\label{defl_angle_gen}
    \hat{\alpha} = 2|\phi| - \pi.
\end{equation}
The analytical integration of Eq.~\eqref{delta_phi_deflection} is infeasible, but in the weak-field regime, following Ref.~\cite{narlikar2010introduction}, we can find the solution in a perturbative manner similar to Sec.~\ref{sec:perihelion}. Introducing the variable $u=1/r$ in Eq.~\eqref{eq_orbit_m0} and calculating the derivative, we obtain
\begin{equation}
    \frac{d^2u}{d\phi^2} + u = \frac{3mu^2}{(1+l)^3}\left(1-\frac{lu}{1+lu}\right),
\end{equation}
or, in approximate form,
\begin{equation}\label{eq_orbit_approx_FW}
    \frac{d^2u}{d\phi^2} + u \approx 3mu^2(1-4lu).
\end{equation}
Firstly, we consider the no-bending approximation satisfying the equation
\begin{equation}\label{nobending_eq}
    \frac{d^2u}{d\phi^2} + u = 0,
\end{equation}
with the general solution
\begin{equation}\label{nobending_sol}
    u = \frac{\cos \phi}{R_0}.
\end{equation}
In the next order of approximation, we substitute Eq.~\eqref{nobending_sol} into the right-hand side of Eq.~\eqref{eq_orbit_approx_FW}. Hence, the resulting equation
\begin{equation} \label{bending_eq}
    \frac{d^2u}{d\phi^2} + u = \frac{3m \cos^2 \phi}{R_0^2}\left(1-\frac{4l \cos \phi}{R_0}\right)
\end{equation}
has a solution given by
\begin{eqnarray}\label{bending_sol}
    u &=& \frac{\cos \phi}{R_0} + \frac{m}{2R_0^2}\big(3-\cos2\phi \big)\nonumber\\ &-& \frac{3ml}{8R_0^3}\big(9\cos \phi - \cos3\phi + 12\phi\sin \phi\big).
\end{eqnarray}
Then, we impose $u=0$ (since particles come from infinity $r\rightarrow\infty$) and substitute Eq.~\eqref{defl_angle_gen} in \eqref{bending_sol}. The resulting expression for the total bending angle takes the form:
\begin{equation}\label{defl_angle_FW}
    \hat{\alpha} = \frac{4m}{R_0}\left(1-\frac{9\pi l}{8R_0}\right).
\end{equation}
It should be noted that the limiting case $l\rightarrow 0$ corresponds to the Schwarzschild spacetime for which $\hat{\alpha} = \frac{4m}{R_0}$. Hence, the effect of the magnetic charge appears at the order $\sim 1/R_0^2$.


\subsection{Deflection of photons} \label{sec:defl_photon}

To calculate the deflection angle of photons, we follow the same approach as in Sec.~\ref{sec:defl_massless}. From the equations of motion \eqref{eq_phi_phot}-\eqref{eq_r_phot}, we derive the following equation for the orbit:
\begin{equation}\label{eq_orbit_phot}
    \frac{1}{r^4}\left(\frac{dr}{d\phi}\right)^2 = \frac{\mathcal{L}_{\mathcal{F}}^2 r^4}{\Phi^2}\left[\frac{1}{\tilde{b}^2} - \frac{\Phi}{r^2 \mathcal{L}_{\mathcal{F}}}\left(1-\frac{2M(r)}{r}\right)\right],
\end{equation}
which yields
\begin{equation}\label{delta_phi_defl_phot}
    \phi = \pm \int_{R_0}^\infty \frac{\Phi}{r^2 \mathcal{L}_{\mathcal{F}}}\left[\frac{1}{\tilde{b}^2} - \frac{\Phi}{r^2 \mathcal{L}_{\mathcal{F}}}\left(1-\frac{2M(r)}{r}\right)\right]^{-1/2}dr,
\end{equation}
whose solution can  be found only by numerical methods. Unsing the coordinate $u=1/r$ in the weak-field regime, we obtain the following second-order differential equation:
\begin{equation}\label{eq_orbit_approx_phot}
    \frac{d^2u}{du^2} + u \approx 3mu^2\left[1+l\left(2mu+\frac{15}{4}\right)\right].
\end{equation}
By inserting the no-bending limit solution Eq.~\eqref{nobending_sol} into Eq.~\eqref{eq_orbit_approx_phot}, we get:
\begin{equation}
    \frac{d^2u}{du^2} + u = \frac{3m \cos^2 \phi}{R_0^2}\left[1+l\left(\frac{2m \cos\phi}{R_0}+\frac{15}{4}\right)\right],
\end{equation}
which leads to the following solution for the deflected photon trajectory:
\begin{eqnarray}\label{bending_phot_sol}
    u &=& \frac{\cos \phi}{R_0} + \frac{m}{2R_0^2}\big(3-\cos2\phi \big) - \frac{ml}{16R_0^3}\Bigg(\frac{30R_0}{m} +9\cos\phi\nonumber\\ &-& \frac{10R_0}{m}\cos 2\phi - \cos 3\phi + 12\phi\sin\phi\Bigg).
\end{eqnarray}
Using Eq.~\eqref{defl_angle_gen}, we finally obtain the deflection angle of photons:
\begin{equation}\label{defl_angle_phot}
    \hat{\alpha} = \frac{4m}{R_0}- \frac{l}{R_0} \left(5+\frac{3\pi m}{4R_0}\right).
\end{equation}
Eq.~\eqref{defl_angle_phot} in physical units takes the following form:
\begin{equation}\label{defl_angle_phot_phys}
    \hat{\alpha} = \frac{4Gm}{R_0 c^2}\left[1-\frac{c^2 l}{4Gm}\left(5+\frac{3\pi G m}{4R_0 c^2}\right)\right].
\end{equation}

\begin{figure*}[ht]
\center{\includegraphics[width=0.31\linewidth]{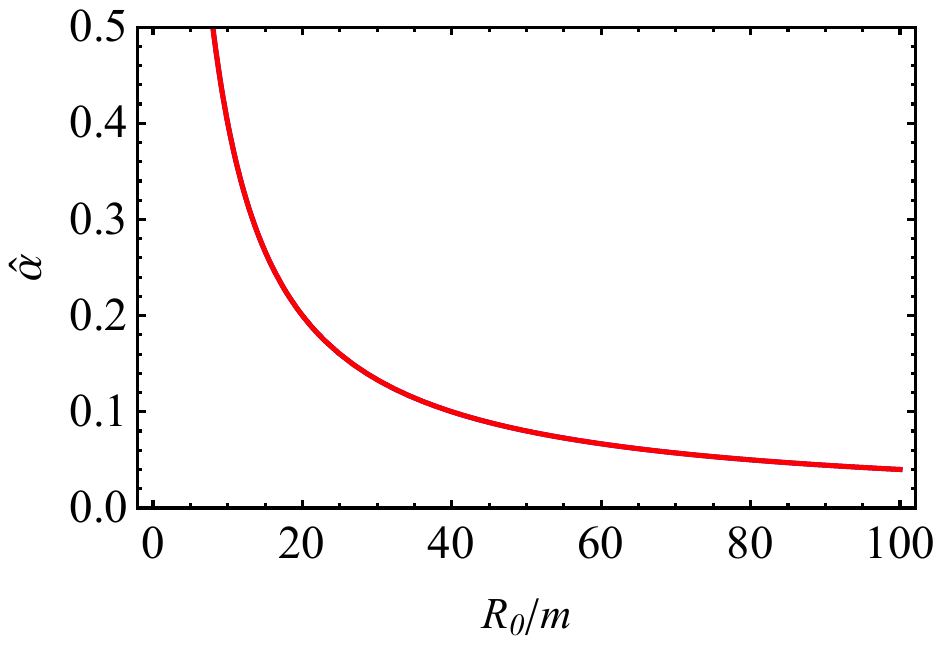}\quad \includegraphics[width=0.31\linewidth]{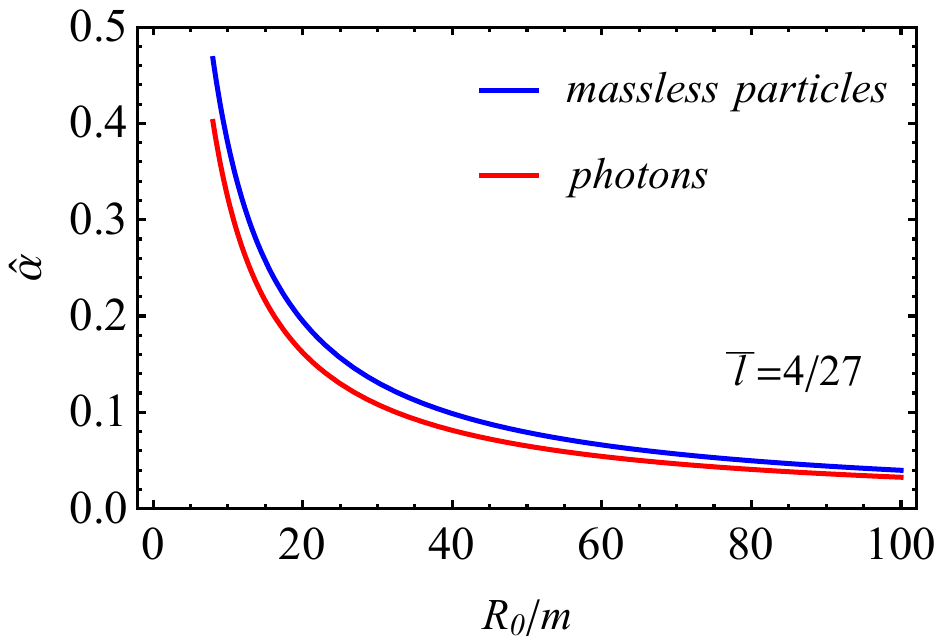}\quad\includegraphics[width=0.31\linewidth]{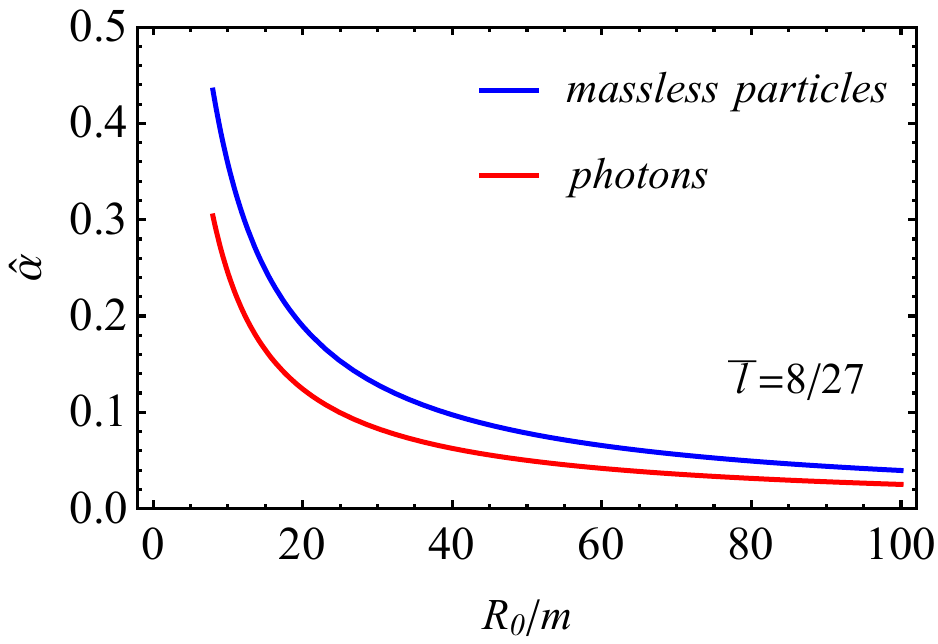}}
\caption{Deflection angles $\hat{\alpha}$ for photons and massless particles versus the normalized distance of closest approach $R_0/m$. Left panel: $\bar{l}=l/m=0$, Middle panel: $\bar{l}=l/m=4/27$, Right panel: $\bar{l}=l/m=8/27$.}
\label{fig:defl_combined}
\end{figure*}

In Fig.~\ref{fig:defl_combined},  we illustrate the dependence of the deflection angle $\hat{\alpha}$ on the closest approach distance, normalized by the mass parameter $R_0/m$, for both photons and massless particles. In the zero charge limit, both cases converge to the classical result $\hat{\alpha} = 4m/R_0$. It can be seen that an increase in the charge parameter $l$ reduces the deflection angle for both photons and massless particles. Additionally, the deflection angle for massless particles is slightly larger than that for photons. This distinction becomes more pronounced as the charge increases, indicating that photons are less affected by the charge parameter than massless particles. It is worth mentioning that a similar trend was obtained in \cite{2019ApJ...887..145S} where the authors found a numerical solution to the deflection angles of massless particles and photons in the Fan-Wang spacetime.

\section{Gravitational Redshift}\label{sec:grav_redshift}

The phenomenon of gravitational redshift occurs when light is observed to be emitted from a region with strong gravitational fields. We will use the standard geometric procedure for defining the redshift with the  effective metric \eqref{eff_metr},  instead of the usual one given in Eq.~\eqref{metr_generic}. In doing so, we consider a photon emitted from a radial position $r_{\text{source}}$ and observed at $r_{\text{obs}} \rightarrow \infty$.

The time component of the effective Fan-Wang metric $\tilde{g}_{00}(r)$ governs the gravitational redshift experienced by the photon as it escapes from the field of the BH. The redshift $z$ is given by:
\begin{eqnarray}
        1+z &=& \sqrt{\frac{-\tilde{g}_{00}(r_{\text{obs}})}{-\tilde{g}_{00}(r_{\text{source}})}}=\frac{1}{\sqrt{-\tilde{g}_{00}(r_{\text{source}})}}\nonumber\\&=& \left[\frac{1}{\mathcal{L}_{\mathcal{F}}}\left(1-\frac{2M(r_{\text{source}})}{r_{\text{source}}}\right)\right]^{-\frac{1}{2}}.
\end{eqnarray}
\begin{figure}[ht]
\center
\includegraphics[width=0.9 \linewidth]{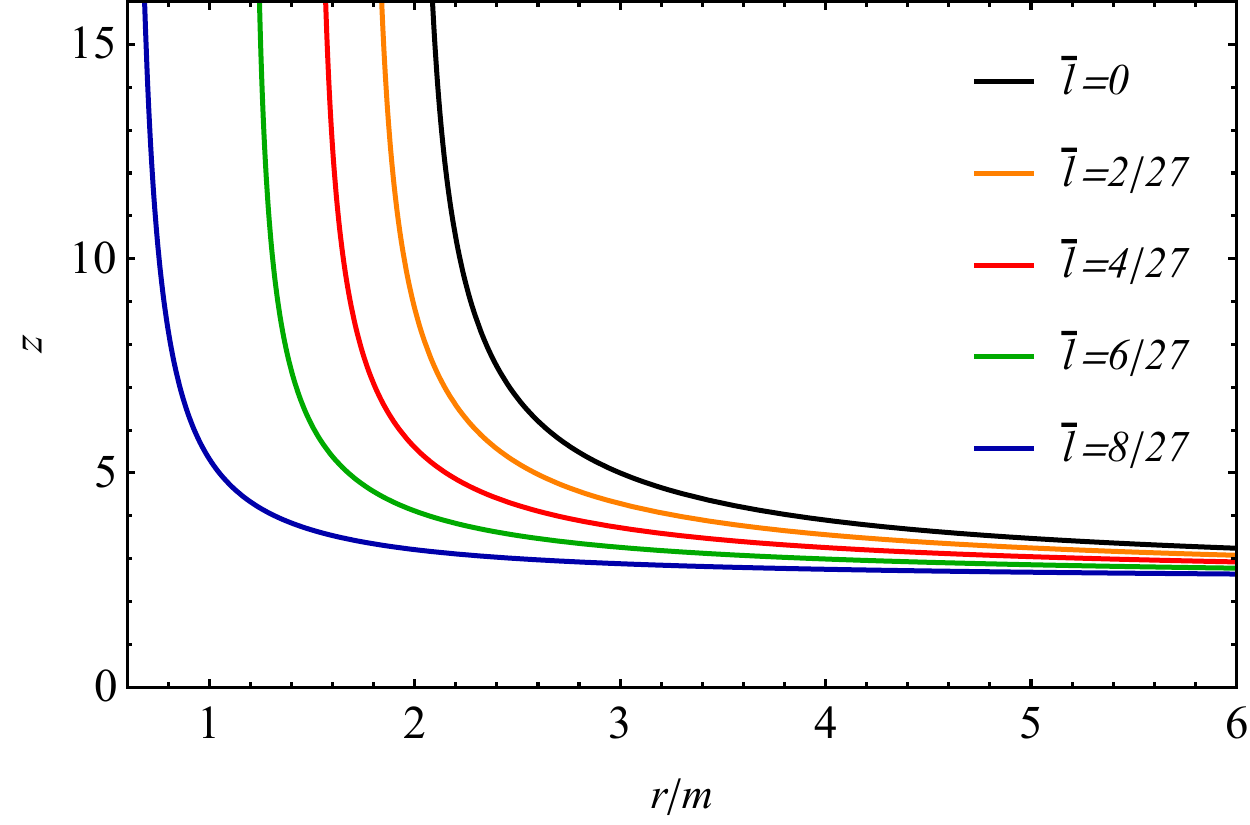}
\caption{Redshift $z$ as a function of normalized distance $r/m$ at different values of charge parameter $\bar{l}=l/m$.}
\label{fig:redshift_fig}
\end{figure}
Figure \ref{fig:redshift_fig} shows the gravitational redshift dependence on the radial coordinate of the source for different values of the charge parameter. As $r$ decreases towards the event horizon, located at the distance where $g_{00}(r_H)=0$, the redshift $z$ increases. Additionally, for the charge parameter increasing to its critical value, $z$ becomes smaller. This behavior is characteristic of strong gravitational fields and the NED effects present in the vicinity of BH.

\section{Final remarks and perspectives} \label{sec:concl}

Here, we studied a particular RBH  obtained from NED and described by the Fan-Wang metric. In particular, we analyzed, in the framework of general relativity, the main features of the model as it appears to be particularly simple to exhibit measurable corrections to the standard Schwarzschild solution. Hence, we motivated our analyses arguing that the presence of a magnetic charge in the Fan-Wang metric may lead to measurable effects. Indeed, our results show that this is possible. 

We focus on the analysis of the gravitational capture, perihelion shift, gravitational deflection, and redshift by examining the motion of massive and massless particles, including photons, on the equatorial plane of the Fan-Wang spacetime. Consequently, we investigate the net NED effects through the deviations from the Schwarzschild spacetime, as remarked above.

For massive particles, we observed that as the charge parameter approaches its critical value, the probability of gravitational capture decreases significantly. This effect could influence the dynamics around charged BHs, particularly in high-velocity environments such as galactic centers or active galactic nuclei. Similarly, for massless particles and photons, the capture cross-section decreases with increasing charge, with notable distinctions between the two cases becoming more pronounced at higher charge values.

The above effects seem to work against an efficient accretion in BHs. Somehow, if the charge has been always non-negligible in each BH history, in the case of supermassive BHs, their formation is more challenging than previously thought. Since the early growth of BHs is largely driven by accretion, a suppression of gravitational capture due to increasing charge would imply longer timescales for mass accumulation.

Moreover, our analysis of the perihelion shift revealed that the charge parameter reduces the magnitude of the shift, with more pronounced effects for orbits closer to the source and those with higher eccentricities. In astrophysical systems, such as the orbits of stars around supermassive black holes, the Fan-Wang corrections introduced slight deviations from classical predictions, \emph{suggesting a potential observational signature of the aforementioned magnetic charge}. Nevertheless, while magnetic charges have not yet been detected, their theoretical investigation remains an active area of research.

Regarding gravitational deflection, our results indicated that an increasing charge leads to a reduction in the deflection angle for both massless particles and photons, with massless particles exhibiting slightly stronger deflection than photons. This effect could have implications for gravitational lensing studies, where different types of radiation (e.g., electromagnetic waves versus neutrino-like particles) may experience varying degrees of bending. 
Remarkably, \emph{such distinctions could serve as potential probes of the underlying properties of BHs}.

Our analysis of the gravitational redshift exhibited a decrease in the observed redshift as the charge parameter increases, indicating a weaker gravitational field near the horizon compared to a Schwarzschild BH of the same mass. 
These findings can be further supported by examining the curvature invariants and energy conditions. Notably, the reduction of the Kretschmann scalar with increasing charge correlates with the observed decreases in all the studied gravitational effects.

Future research could extend these findings in several manners. One promising avenue is to analyze the rotating counterpart of the Fan-Wang solution, investigating how the angular momentum of the source influences gravitational phenomena. This may be worked out by using the Newman-Janis algorithm, for example. Additionally, numerical simulations in realistic astrophysical environments, such as active galactic nuclei or regions with strong magnetic fields, could bridge theoretical predictions with observations. Furthermore, we may explore the implications of NED even in other contexts and/or metrics. Last but not least, classical effects may even be put in correspondence with semiclassical results \cite{2014PhRvD..90h4032L,2021PhRvD.104j5020G,2024CQGra..41l5011L}, with the intent to find out a connection between infrared and ultraviolet regimes of gravity.\vspace{3mm}

\section{Acknowledgements}
YeK acknowledges Grant No. AP23488743, TK acknowledges Grant No. AP19174979, GS acknowledges Grant No. AP19575366, AU acknowledges Grant No. AP19680128 and KB acknowledges Grant No. BR21881941 all from the Science Committee of the Ministry of Science and Higher Education of the Republic of Kazakhstan. OL acknowledges financial support from the Fondazione ICSC, Spoke 3 Astrophysics and Cosmos Observations, National Recovery and Resilience Plan (Piano Nazionale di Ripresa e Resilienza, PNRR) Project ID CN$\_$00000013 "Italian Research Center on High-Performance Computing, Big Data and Quantum Computing" funded by MUR Missione 4 Componente 2 Investimento 1.4: Potenziamento strutture di ricerca e creazione di "campioni nazionali di R$\&$S (M4C2-19 )" - Next Generation EU (NGEU) GRAB-IT Project, PNRR Cascade Funding Call, Spoke 3, INAF Italian National Institute for Astrophysics, Project code CN00000013, Project Code (CUP): C53C22000350006, cost center STI442016. HQ acknowledges the support of UNAM-DGAPA-PAPIIT, Grant No. IN108225.

\bibliography{0refs}
\end{document}